\documentclass[twocolumn,           
               showpacs,            
               preprintnumbers,     
               aps,                 
               prd,                 
               a4paper,             
               superscriptaddress,  
               nofootinbib,         
               tightenlines,        
               floats,floatfix      
               ]{revtex4-1}

\pdfoutput=1

\usepackage{ graphicx }
\usepackage{ latexsym }
\usepackage{ amsmath, amssymb }        
\usepackage[ draft=false ]{ hyperref }
\usepackage{multirow}

\newcommand{\rd}{\mathrm{d}}

\newcommand{\R}{\mathbb{R}}
\newcommand{\e}{\mathrm{e}}

\newcommand{\h}{\mathrm{h}}

\begin{document}


\title{Characterization  of the gravitational  wave emission  of three
  black            holes}            \author{Pablo            Galaviz}
\email{Pablo.Galaviz@monash.edu}   

\affiliation{School   of  Mathematical  Science,   Monash  University,
  Melbourne, VIC 3800, Australia}
\affiliation{Theoretical  Physics
  Institute, University of Jena, 07743 Jena, Germany}

    \author{Bernd      Br\"ugmann}
\email{Bernd.Bruegmann@uni-jena.de}  \affiliation{Theoretical  Physics
  Institute, University of Jena, 07743 Jena, Germany}


\date{\today}


\begin{abstract}
  We study  the gravitational wave  emission of three  compact objects
  using  post-Newtonian  (PN) equations  of  motion  derived from  the
  Arnowitt-Deser-Misner Hamiltonian formulation, where we include (for
  the  first time  in  this context)  terms  up to  2.5~PN order.   We
  perform  numerical simulations  of a  hierarchical  configuration of
  three  compact bodies in  which a  binary system  is perturbed  by a
  third,  lighter  body  initially  far  away from  the  binary.   The
  relative importance of the different PN orders is examined.
  We  compute  the waveform  in  the  linear  regime considering  mass
  quadrupole, current quadrupole and mass octupole contributions.
  Performing a  spherical harmonic  decomposition of the  waveforms we
  find that from the $l=3$ modes it is possible to extract information
  about the third body, in  particular, the period, eccentricity of its
  orbit, and the inclination angle  between the inner and outer binary
  orbits.
\end{abstract}




\maketitle




\section{Introduction}
\label{sec:introduction}

In the near future gravitational wave detectors will open a new branch
on  astronomy  beyond electromagnetism  and  particles.   In order  to
extract  astrophysical information  from the  waves it  is  crucial to
model  the  source in  an  accurate way.   One  of  the parameters  to
determine is the number of bodies which generate the waves.
Some configurations  of three  bodies can produce  particular periodic
waveforms with distinctive features, e.g.\ \cite{ChiImaAsa07}.  On the
other  hand,  Lagrange's   triangle  solution  produces  a  quadrupole
waveform which  is identical  to the one  produced by a  binary system
\cite{TorHatAsa09}.  However, it is  possible to distinguish between a
binary system and a triple one by considering the octupole part of the
waveform \cite{Asa09}.
Lagrange's solution  is stable  only if one  of the bodies  holds more
than 95\% of the total mass \cite{ValKar06}. For three-body systems of
comparable  masses  there are  stable  configuration  in  which it  is
possible to  characterize the  system by looking  at the  waveform. In
this  work  we  consider   \textit{Jacobian  systems},  in  which  the
three-body configuration  is composed of two parts,  a clearly defined
binary and a third body orbiting far away.  We will refer to this kind
of system also as a hierarchical system.

Several models of three or more black holes were recently studied from
the astrophysical point of view.
Hierarchical three black hole configurations interacting in a galactic
core   were   studied   by    several   authors.    For   example   in
\cite{GulMilHam03,GulMilHam04,GulMilHam05a}   some  configurations  of
intermediate-mass  black   holes  with  different   mass  ratios  were
considered.  The inclusion of  gravitational radiation was done via an
effective  force  which  includes  1~PN  (post-Newtonian)  and  2.5~PN
corrections to  the binary dynamics.  The configurations  consist of a
binary  system  in  a  quasicircular  orbit and  a  third  black  hole
approaching from  a distance around  200 times the  binary separation.
The initial eccentricity was specified in a random way.
N-body  simulations  of   dynamical  evolution  of  triple  equal-mass
supermassive  black  holes in  a  galactic  nuclei  were performed  in
\cite{IwaFunMak06}.   The  method  includes  an effective  force  with
gravitational radiation terms and galaxy halo interactions.
In \cite{HofLoe07} the dynamics  of repeated triple supermassive black
hole  interactions in  galactic nuclei  with several  mass  ratios and
eccentricities were considered.   The simulations were performed using
Newtonian dynamics with corrections  through an additional force which
includes  2.5~PN  corrections  to  the  binary  dynamics  and  stellar
dynamical friction.
Other  astrophysical applications of  multiple black  hole simulations
include,       for       example,      three-body       kicks
\cite{Hofloe06,GuaPorSip05}  and binary-binary  encounters  (see e.g.\
\cite{Mik83,Mik84,MilHam02,Hei01,ValMik91}).

The  first complete  simulations using  general-relativistic numerical
evolutions    of    three    black    holes    were    presented    in
\cite{CamLouZlo07f,LouZlo07a}.  These recent simulations show that the
dynamics of  three compact  objects display a  qualitatively different
behavior than the Newtonian dynamics.
In   \cite{GalBruCao10a}  the   sensitivity   of  fully   relativistic
evolutions of  three and  four black holes  to changes in  the initial
data was examined,  where the examples for three  black holes are some
of      the      simpler      cases     already      discussed      in
\cite{CamLouZlo07f,LouZlo07a}.
The apparent horizon and the event horizon of multiple black holes have
been  studied in  \cite{Die03a,JarLou10,PonLouZlo10}.   Although fully
general-relativistic  simulations are  available, they are  limited to
only  a small  number of  orbits for  small separations  of  the black
holes.

In the present work we study three-body systems with PN methods, where
the main technical novelty is the inclusion of the 2.5 PN terms in the
orbital  dynamics.  We  do  not consider  compact  objects with  spin,
although  recently  the knowledge  of  Hamiltonian  up  to 2.5~PN  was
completed with  the computation of a  next-to-leading order spin-orbit
and spin-spin Hamiltonian \cite{DamJarSch07a,SteHerSch08,HarSte10}

Using post-Newtonian techniques, it  is currently possible to describe
the dynamics  of $n$ compact objects  without spin up  to 3.5~PN order
(see    e.g.\   \cite{JarSch97,    Bla06a,    FutIto07,   DamJarSch00,
  DamJarSch00a,  DamJarSch01,  BlaDamEsp04a}),  although explicit  and
closed  expressions for  the terms  required for  our purpose  are not
available    for   arbitrary    $n$.    For    binary    systems   the
Arnowitt-Deser-Misner  (ADM) Hamiltonian  has been  specialized  up to
3.5~PN order \cite{KonFaySch03}.  For  three bodies there are explicit
formulas up to 2.5~PN order,  with the final key integral performed in
\cite{Sch87}   (see   also  Appendix~\ref{sec:first-second-post}   and
\cite{LouNak08}; see \cite{Chu09}  for an alternative discussion where
the final result is not as yet explicit).
For four compact objects, the same degree of explicitness has not been
obtained, see  e.g.~\cite{OhtKimHii75} on which  \cite{Sch87} is based
where the  integral (C.2) must be  computed for four  bodies, and also
Appendix~D of  \cite{MitWil07}.  In the  ADM-Hamiltonian formalism the
resulting  equations  of  motion  exactly conserve  the  constants  of
motion. For  numerical simulations  this represents an  advantage with
respect  to other  post-Newtonian approaches,  since the  constants of
motion can be tracked and their constancy continually checked.

Periodic solutions were studied  using the 1~PN and 2~PN approximation
in  \cite{Moo93,TatTakHid07,LouNak08}.    Examples  of  three  compact
bodies   in    a   collinear   configuration    were   considered   in
\cite{YamAsa10,YamAsa10a},   and  Lagrange's   equilateral  triangular
solution was studied including 1~PN effects in \cite{IchYamAsa10}.  In
\cite{JSch10}, the stability of the  Lagrangian points in a black hole
binary  system was  studied  in  the test  particle  limit, for  which
radiation effects were modeled by a drag force.

The  most likely  source  of gravitational  waves  are binary  compact
objects.  Recently  it was shown  that the probability that  more than
two black holes  interact in the strongly relativistic  regime is, not
surprisingly, very small  \cite{AmaDew10}.  For practical purposes the
creation  of gravitational waveform  templates for  gravitational wave
detectors  is naturally  focused on  binary systems,  and  even binary
systems  can produce  complicated waveforms  when taking  into account
spinning black holes and eccentric orbits, e.g.\ \cite{LevCon10}.

Nevertheless,  it remains  an interesting  question of  principle what
additional  wave phenomena  are  possible for  more  than two  compact
objects.
The  waveform characterization  of three  or more  compact  objects is
complementary to the study of binary systems.  For a three-body system
the complexity of the orbits  can reveal properties of the waves which
for  a  binary  system are  hidden.   As  we  will demonstrate  for  a
hierarchical system, from the $l=3$ modes of the gravitational wave it
is possible to extract  information about the third body, particularly
the  period,  eccentricity of  its  orbit  and  the inclination  angle
between the inner and outer binary orbits.

The paper is organized as follows. In Sec.~\ref{sec:equations-motion},
we  summarize  the  equation   of  motion  up  to  2.5  post-Newtonian
approximation for  three bodies. This  is followed by a  discussion of
gravitational  radiation  in  the  linear  regime  and  the  multipole
expansion of the gravitational waves.
In Sec.~\ref{sec:numer-integr},  we describe the  numerical techniques
used to solve  the equation of motion and we  present some results for
test cases.
The perturbation of a binary system  by a third object is presented in
Sec.~\ref{sec:strong-pert-binary},   where    we   perform   numerical
experiments in order to characterize the waveform.
We conclude in Sec.~\ref{sec:discussion}.

\subsection{Notation and units}
\label{sec:notation-units}

We employ  the following notation: $\vec{x}=(x_i)$ denotes  a point in
the three-dimensional Euclidean  space $\R^3$, letters $a,b,\dots$ are
particle    labels.     We   define    $\vec{r}_a:=\vec{x}-\vec{x}_a$,
$r_a:=\vert \vec{r}_a \vert$,  $\hat{n}_a:=\vec{r}_a/r_a$; for $a \neq
b$,  $\vec{r}_{ab}:=\vec{x}_a-\vec{x}_b$,  $r_{ab}:=\vert \vec{r}_{ab}
\vert$, $\hat{n}_{ab}:=\vec{r}_{ab}/r_{ab}$;  here $\vert \cdot \vert$
denotes  the length  of a  vector. The  mass parameter  of  the $a$-th
particle  is denoted by  $m_a$, with  $M=\sum_a m_a$.   Summation runs
from 1 to  3. The linear momentum vector is  denoted by $\vec{p}_a$. A
dot  over a  symbol, as  in  $\dot{\vec{x}}$, means  the total  time
derivative,  and  partial differentiation  with  respect  to $x^i$  is
denoted by $\partial_i$.

In  order  to  simplify  the  calculations  it  is  useful  to  define
dimensionless  variables (see  e.g.\ \cite{Szi06}).   We use  as basis
quantities  for  the  Newtonian  and  post-Newtonian  calculation  the
gravitational constant $G$, the speed  of light $c$ and the total mass
of the system $M$. Using derived constants for time $\tau = M G /c^3$,
length $\mathit{l}=M  G /c^2$, linear momentum $\mathcal{P}=  M c$ and
energy $\mathcal{E}= M c^2$  we construct dimensionless variables. The
physical  variables are  related with  the dimensionless  variables by
means  of a  scaling, for  example, denoting  with capital  letters the
physical variables  with the usual  dimensions and with  lowercase the
dimensionless  variable we  define  for a  particle  $a$ its  position
$\vec{x}_a:=\vec{X}_a      /     \mathit{l}$,      linear     momentum
$\vec{p}_a:=\vec{P}_a / \mathcal{P}$ and mass $m_a=M_a/M$ (notice that
$m_a<1$, $\forall a$).

\section{Equations of motion}
\label{sec:equations-motion}

In  the  ADM post-Newtonian  approach  it  is  possible to  split  the
Hamiltonian in a series with  coefficients which are inverse powers of
the speed of light (see e.g.\ \cite{Bla06a,Mag07a})
\begin{equation}\label{eq:1}
\mathcal{H}_{\leq 2.5} = \mathcal{H}_0 + c^{-2}\mathcal{H}_{1} + 
  c^{-4}\mathcal{H}_{2} + c^{-5}\mathcal{H}_{2.5}.
\end{equation}
Here  each  term  of  the Hamiltonian  $c^{n}\mathcal{H}_{n/2}$  is  a
quantity with a dimension of  energy, and we  write it  explicitly with
factors   of  $c$.    The  dimensionless   Hamiltonian  is   given  by
$H_{n/2}=c^{n}\mathcal{H}_{n/2}  /  \mathcal{E}$.   For each  term  we
calculate the equations of motion
\begin{eqnarray}
(\dot{x}^i_a)_{n} &=& \frac{\partial H_{n}}{\partial p^i_a},
\label{eq:2} \\
 -(\dot{p}^i_a)_{n}  &  = &  \frac{\partial H_{n}}{\partial x^i_a}, 
\label{eq:3}
\end{eqnarray}  
where the equations of motion up to 2.5~PN approximation are
\begin{eqnarray}
\dot{\vec{x}}_a &=& (\dot{\vec{x}}_a)_{0}+(\dot{\vec{x}}_a)_{1}+
(\dot{\vec{x}}_a)_{2}+(\dot{\vec{x}}_a)_{2.5},\label{eq:4}\\ 
\dot{\vec{p}}_a &=& (\dot{\vec{p}}_a)_{0}+(\dot{\vec{p}}_a)_{1}+
(\dot{\vec{p}}_a)_{2}+(\dot{\vec{p}}_a)_{2.5}.\label{eq:5}
\end{eqnarray}  

The first  term in \eqref{eq:1}  is the Hamiltonian for  $n$ particles
interacting under Newtonian gravity,
\begin{equation}\label{eq:6}
H_0    =    \frac{1}{2}    \sum^n_a   \frac{\vec{p}_a^{\;2}}{m_a}    -
\frac{1}{2}\sum^n_{a,b \neq a} \frac{m_a m_b}{r_{ab}},
\end{equation}
with   $\vec{p}_a=m_a   \dot{\vec{x}}_a^{\;2}$.    The  inclusion   of
post-Newtonian corrections enriches the phenomenology of the system.

\subsubsection{Post-Newtonian equations of motion up to 2.5 order} 
\label{sec:post-newt-equat}

The  first post-Newtonian  correction to  the equations  of  motion is
discussed      extensively       in      the      literature      (see
e.g.\ \cite{DamSch85,Bla06a}).
The three-body Hamiltonian at first and second post-Newtonian order is
given in Appendix~\ref{sec:first-second-post}.
The equations of  motion for the first post-Newtonian  order are given
by \eqref{eq:2}, \eqref{eq:3} and  \eqref{eq:53}.  For particle $a$ we
obtain
\begin{widetext}
\begin{equation}\label{eq:7}
(\dot{\vec{x}}_a)_{1} = -\frac{\vec{p}_a^{\;2}}{2 m_a^3}\vec{p}_a -
\frac{1}{2} \sum_{b \neq a} \frac{1}{r_{ab}} \left ( 6 \frac{m_b}{m_a} 
\vec{p}_a - 7 \vec{p}_b- (\hat{n}_{ab} \cdot \vec{p}_b ) \hat{n}_{ab} \right ),
\end{equation}
\begin{equation}\label{eq:8}
\begin{split}
(\dot{\vec{p}}_a)_{1} =& -\frac{1}{2}\sum_{b\neq a} \left [ 
3\frac{m_b}{m_a} \vec{p}_a^{\;2} + 3 \frac{m_a}{m_b} \vec{p}_b^{\;2}- 
7(\vec{p}_a \cdot \vec{p}_b) -3(\hat{n}_{ab} \cdot \vec{p}_a) (\hat{n}_{ab} 
\cdot \vec{p}_a)  \right]\frac{\hat{n}_{ab}}{r_{ab}^2} \\&+ \sum_{b \neq a}
\sum_{c \neq a} \frac{m_a m_b m_c}{r_{ab}^2 r_{ac}} \hat{n}_{ab} + \sum_{b \neq a}
\sum_{c \neq b} \frac{m_a m_b m_c}{r_{ab}^2 r_{bc}} \hat{n}_{ab}
- \frac{1}{2}
\sum_{a \ne b} \left[ \frac{(\hat{n}_{ab} \cdot \vec{p}_b )\vec{p}_a  +  
(\hat{n}_{ab} \cdot \vec{p}_a)\vec{p}_b  }{r_{ab}^2} \right].
\end{split}
\end{equation}
For the  second post-Newtonian  approximation the equations  of motion
are  calculated using  \eqref{eq:2},  \eqref{eq:3} and  \eqref{eq:54}.
For brevity we do not display the explicit equations.

Following  \cite{JarSch97,KonFaySch03} we  obtain equations  of motion
from the 2.5~PN Hamiltonian in the ADM gauge.
The general 2.5~PN Hamiltonian is
\begin{equation}\label{eq:9}
H_{2.5} = \frac{1}{45} \dot{\chi}_{(4)ij}(\vec{x}_{a'},\vec{p}_{a'};t) 
\chi_{(4)ij}(\vec{x}_{a},\vec{p}_{a}),
\end{equation}
where the auxiliary function $\chi_{(4)ij}$ is defined by
\begin{equation}\label{eq:10}
\chi_{(4)ij}(\vec{x}_{a},\vec{p}_{a}) := \sum_a 
\frac{2}{m_a} \left( \vec{p}_a^{\;2} \delta_{ij} - 3 p_{ai} p_{aj} 
\right)  + \sum_{a} \sum_{b\neq a}
\frac{m_a m_b}{r_{ab}} \left( 3 n_{abi}n_{abj}  - \delta_{ij} 
\right).
\end{equation}
Our  expressions  differ  from  \cite{JarSch97,KonFaySch03} due  to  a
different choice  of units.   The explicit form  of the  derivative in
\eqref{eq:9} is
\begin{equation}\label{eq:11}
\begin{split}
\dot{\chi}_{(4)ij}(\vec{x}_{a'},\vec{p}_{a'}) = &  \sum_{a'} 
\frac{2}{m_{a'}} \left[ 2 (\dot{\vec{p}}_{a'} \cdot \vec{p}_{a'}) \delta_{ij} - 
3 (\dot{p}_{a'i} p_{a'j}+p_{a'i} \dot{p}_{a'j}) \right]    
 \\  +& \sum_{a'} \sum_{b'\neq a'}
\frac{m_{a'} m_{b'}}{r_{a'b'}^2} \left[ 3 (\dot{r}_{a'b'i}n_{a'b'j} +  n_{a'b'i} \dot{r}_{a'b'j}) \right.
 + \left. (\hat{n}_{a'b'}\cdot \dot{\vec{r}}_{a'b'}) (\delta_{ij}-9 n_{a'b'i} n_{a'b'j})  \right].
\end{split}
\end{equation}
We denote  the retarded variables by primed  quantities.  The position
and momentum  appearing in Eq.~\eqref{eq:11}  are not affected  by the
derivative operators given by  \eqref{eq:2} and \eqref{eq:3}, and only
after calculating those derivatives  we identify positions and momenta
inside and outside the transverse-traceless variables (i.e.~the primed
and unprimed quantities).
We  replace  the  time  derivatives  of  the  primed  coordinates  and
positions given  in Eq.~\eqref{eq:11} by the 1~PN  equations of motion
Eqs.~\eqref{eq:7} and \eqref{eq:8}.

The equations of motion for 2.5~PN are given in short hand by
\begin{eqnarray}
(\dot{\vec{x}}_a)_{2.5} &=& 
\frac{1}{45} \dot{\chi}_{(4)ij}(\vec{x}_a,\vec{p}_a;(\dot{\vec{x}}_a)_{1},
(\dot{\vec{p}}_a)_{1},t) 
\frac{\partial }{\partial \vec{p}_a} \chi_{(4)ij}(\vec{x}_{a},\vec{p}_{a}),
\label{eq:12}\\
(\dot{\vec{p}}_a)_{2.5} &=& 
-\frac{1}{45} \dot{\chi}_{(4)ij}(\vec{x}_a,\vec{p}_a;(\dot{\vec{x}}_a)_{1},
(\dot{\vec{p}}_a)_{1},t) 
\frac{\partial }{\partial \vec{x}_a} \chi_{(4)ij}(\vec{x}_{a},\vec{p}_{a}).
\label{eq:13}
\end{eqnarray}
\end{widetext}

Given initial values for  $\vec{x}_a$ and $\vec{p}_a$ of each particle
it  is  possible  to  integrate  the  resulting  equations  of  motion
numerically.

\subsection{Gravitational radiation in the linear regime} 
\label{sec:grav-radi-line}

We  consider leading  order and  next-to-leading  order gravitational
waves  calculated  using  trajectories  which  contain  post-Newtonian
corrections.   We  compute the  gravitational  waveforms  for a  given
observational direction, and  alternatively we calculate the multipole
decomposition  which  allows  us  to  reconstruct  the  waves  for  an
arbitrary direction.   The inclusion of  post-Newtonian corrections to
the  gravitational waveforms  is a  topic for  future research  in the
three-compact-body problem.

\subsubsection{Quadrupole and octupole formulas}
\label{sec:quadr-octup-form}

Here  we  summarize the  formulas  for  quadrupole  and octupole  mass
radiation  and for  current  quadrupole radiation  (for  a review  see
e.g.\ \cite{Mag07a,FlaHug05}).  The second  and third mass moments are
defined by
\begin{eqnarray}
M^{ij}(t) &=& \int T^{00}(\vec{x},t) x^i x^j \rd^3 x,\label{eq:14} \\ 
M^{ijk}(t) &=& \int T^{00}(\vec{x},t) x^i x^j x^k \rd^3 x.\label{eq:15}
\end{eqnarray}
The second moment of the momentum density is
\begin{equation}\label{eq:16}
P^{i,jk}(t)=\int T^{0i}(\vec{x},t) x^j x^k \rd^3 x. 
\end{equation}
For $n$ point particles
\begin{equation}\label{eq:17}
T^{\mu \nu}(\vec{x},t)  = \sum_a \frac{p_a^\mu  p_a^\nu}{\gamma_a m_a}
\delta^3(\vec{x}-\vec{x}_a(t)),
\end{equation}
where $\gamma_a:=  (1-\vec{p}_a^{\;2})^{-1/2}$ is the  Lorentz factor,
and  $p_a^\mu :=  \gamma_a (m_a,\vec{p}_a)$  is the  four-momentum. In
this case Eqs \eqref{eq:14}-\eqref{eq:16} reduce to
\begin{eqnarray}
M^{ij}(t) &=& \sum_a \gamma_a m_a x_a^i(t) x_a^j(t),\label{eq:18} \\ 
M^{ijk}(t) &=& \sum_a \gamma_a m_a x_a^i(t) x_a^j(t) x_a^k(t),\label{eq:19} \\ 
P^{i,jk}(t) &=& \sum_a p_a^i(t) x_a^j(t) x_a^k(t).\label{eq:20}
\end{eqnarray}
In the following we consider the case where $\vert \vec{p}_a \vert \ll
1$, $\gamma_a \simeq 1$.

The mass quadrupole and octupole moment are given by 
\begin{eqnarray*}
\mathcal{Q}^{ij}(t) &=& M^{ij}-
\frac{1}{3} \delta^{ij}M^{kk},\\
\mathcal{O}^{ijk}(t) &=& M^{ijk}-
\frac{1}{5}( \delta^{ij}M^{llk}
+ \delta^{ik}M^{ljl} + \delta^{jk}M^{ill}),
\end{eqnarray*}
where  repeated  indices  mean  summation  from 1  to  3.  The  current
quadrupole is given by
\begin{equation}\label{eq:21}
\mathcal{C}^{k,lm}(t) =  P^{k,lm}+P^{l,km}-2P^{m,kl}.
\end{equation}

A projection  tensor into  the plane normal  to the direction  of wave
propagation, $\hat{n}=(\sin  \theta \sin \phi, \sin  \theta \cos \phi,
\cos \theta)$, is defined by
\begin{eqnarray}
\mathcal{P}_{ij}&:=&\delta_{ij}-n_i n_j,\label{eq:22}\\
\Lambda_{ijkl}(\hat{n})&:=&\mathcal{P}_{ik} \mathcal{P}_{jl}- \frac{1}{2} \mathcal{P}_{ij} \mathcal{P}_{kl}.\label{eq:23}
\end{eqnarray}
The mass quadrupole and octupole waveforms are given by
\begin{eqnarray} 
h_{ij}^{TT}(\vec{x},t)_{MQ}&=&\frac{2}{r} \Lambda_{ijkl}(\hat{n})
\ddot{\mathcal{Q}}^{kl}(t-r),\label{eq:24}\\
h_{ij}^{TT}(\vec{x},t)_{MO}&=&\frac{2}{3r} \Lambda_{ijkl}(\hat{n})n_m
\dddot{\mathcal{O}}^{klm}(t-r),\label{eq:25}
\end{eqnarray}
and the current quadrupole contribution to the waveform is
\begin{equation}
h_{ij}^{TT}(\vec{x},t)_{CQ}=\frac{4}{3r} \Lambda_{ijkl}(\hat{n})n_m
\ddot{\mathcal{C}}^{k,lm}.\label{eq:26}
\end{equation}  
The total contribution on the waveform is given by
\begin{equation}
\begin{split}
h_{ij}^{TT}(\vec{x},t)=&h_{ij}^{TT}(\vec{x},t)_{MQ}+h_{ij}^{TT}(\vec{x},t)_{CQ}\\
&+h_{ij}^{TT}(\vec{x},t)_{MO} +\dots.\label{eq:27}
\end{split}
\end{equation}  
where  $\dots$ means  additional multipoles.   Assuming that  the wave
propagates  in  the  $\hat{z}$-direction, then  $h_+=h_{11}^{TT}$  and
$h_{\times}=h_{12}^{TT}$.      For      an     arbitrary     direction
$\hat{n}(\theta,\phi)$ we  have to perform  a rotation of the  axes in
order  to  identify  the   polarization  with  the  $h_{11}^{TT}$  and
$h_{12}^{TT}$ components.

We  decompose  $h_+$  and  $h_{\times}$  into  modes  using  spherical
harmonics with spin-weight minus two,
\begin{equation}\label{eq:28}
h_+-i h_{\times} = \sum_{l} \sum_{m=-l}^{l} \,_{-2}Y^l_m(\Theta,\Phi)\mathrm{h}^{m}_{l},
\end{equation}
where  
\begin{equation}\label{eq:29}
_{s}Y^l_m(\Theta,\Phi) := (-1)^s\sqrt{\frac{2 l + 1 }{4\pi}} d^l_{m(-s)}(\Theta) \e^{im \Phi},
\end{equation}
\begin{equation}\label{eq:30}
\begin{split}
d^l_{ms}(\Theta) :=&\sum_{t=C_1}^{C_2}\frac{(-1)^t[(l+m)!(l-m)!(l+s)!(l-s)!]^{1/2}}{(l+m-t)!(l-s-t)!t!(t+s-m)!}\\
&\times (\cos \Theta/2)^{2l+m-s-2t}(\sin \Theta /2 )^{2t+s-m},
\end{split} 
\end{equation}
with    $C_1=\max(0,m-s)$   and   $C_2=\min(l+m,l-s)$.     Using   the
orthonormality of  the spherical harmonics  it is possible  to compute
$\mathrm{h}^l_m$ by
\begin{equation}\label{eq:31}
\mathrm{h}^l_m = \int_{0}^{2\pi} \int_0^{\pi} \,_{-2}\bar{Y}^l_m(\theta,\pi/2-\phi)( h_+-i h_{\times})  \rd \Omega,
\end{equation}
where $\rd \Omega = \sin \theta \rd \theta \rd \phi$. 

\section{Simulations and results} 
\label{sec:simulations-results}


\subsection{Numerical integration} 
\label{sec:numer-integr}

We    solved    the   equations    of    motion   numerically    using
\textsc{Mathematica} 7.0 \cite{Wol08}.  We used the built-in low-level
functions  of  the  \textit{NDSolve}  routine with  a  ``double-step''
method  using as  subalgorithm the  ``explicit midpoint''  method.  We
divided long  simulations into substeps  in order to store  the result
from time  to time and  to avoid saturating  the random-access-memory.
With this approach we can  produce accurate numerical solutions of the
equations   of   motion.   For   our   purpose   the  performance   of
\textsc{Mathematica} solving the ordinary differential equation
system is not an issue (see  the performance and accuracy tests at the
end of this section).

An important issue in the numerical integration of a three-body system
arises when two of the bodies  come very close to each other. Adaptive
step size methods can automatically maintain the necessary accuracy to
properly resolve  the orbits in  the close interaction, but  issues of
efficiency arise. For the Newtonian system a number of techniques have
been developed that address problems with accuracy and efficiency, see
e.g.\   \cite{Bur67,Heg74,MikAar89,MikAar93,MikAar96}  and  references
therein.   For our  PN evolutions,  efficiency was  not an  issue, and
furthermore  the equations  of motion  are not  valid  for arbitrarily
small separation  anyway. What  is of relevance  here is  a convenient
criterion  of when  to stop  the evolution.   We monitor  the absolute
value  of  each  conservative  part of  the  Hamiltonian  \eqref{eq:1}
relative to the sum of the absolute values,
\begin{equation}\label{eq:32}
H^{\%}_i := 100\left (\frac{\vert H_i \vert}{\vert H_0 \vert + \vert H_1 \vert + \vert H_2 \vert} \right ). 
\end{equation}
We   stop  the  simulation   when  the   contribution  of   the  first
post-Newtonian correction is larger than 10\%.

In the remainder of this section we report on several tests that allow
us to estimate the numerical errors.
We  use the Lagrangian  equilateral triangle  solution to  compare the
numerical  with an  analytical solution.  In Lagrange's  solution each
body  is  sitting  in  one  corner of  an  equilateral  triangle  (see
e.g.~\cite{GolPooSaf01}).   We  set  the  side  of  such  triangle  to
$L=1000$, the mass ratio to  1:2:3, and the eccentricity to zero. Then
each body follows  a circular orbit (with different  radii) around the
center  of   mass.   The   solution  in  this   case  is   not  stable
\cite{ValKar06},  however  for  circular  orbits we  can  compute  the
waveforms and compare with the analytical expressions  \cite{Asa09}.

\begin{figure}[btp]
  \centering
  \includegraphics[width=85mm]{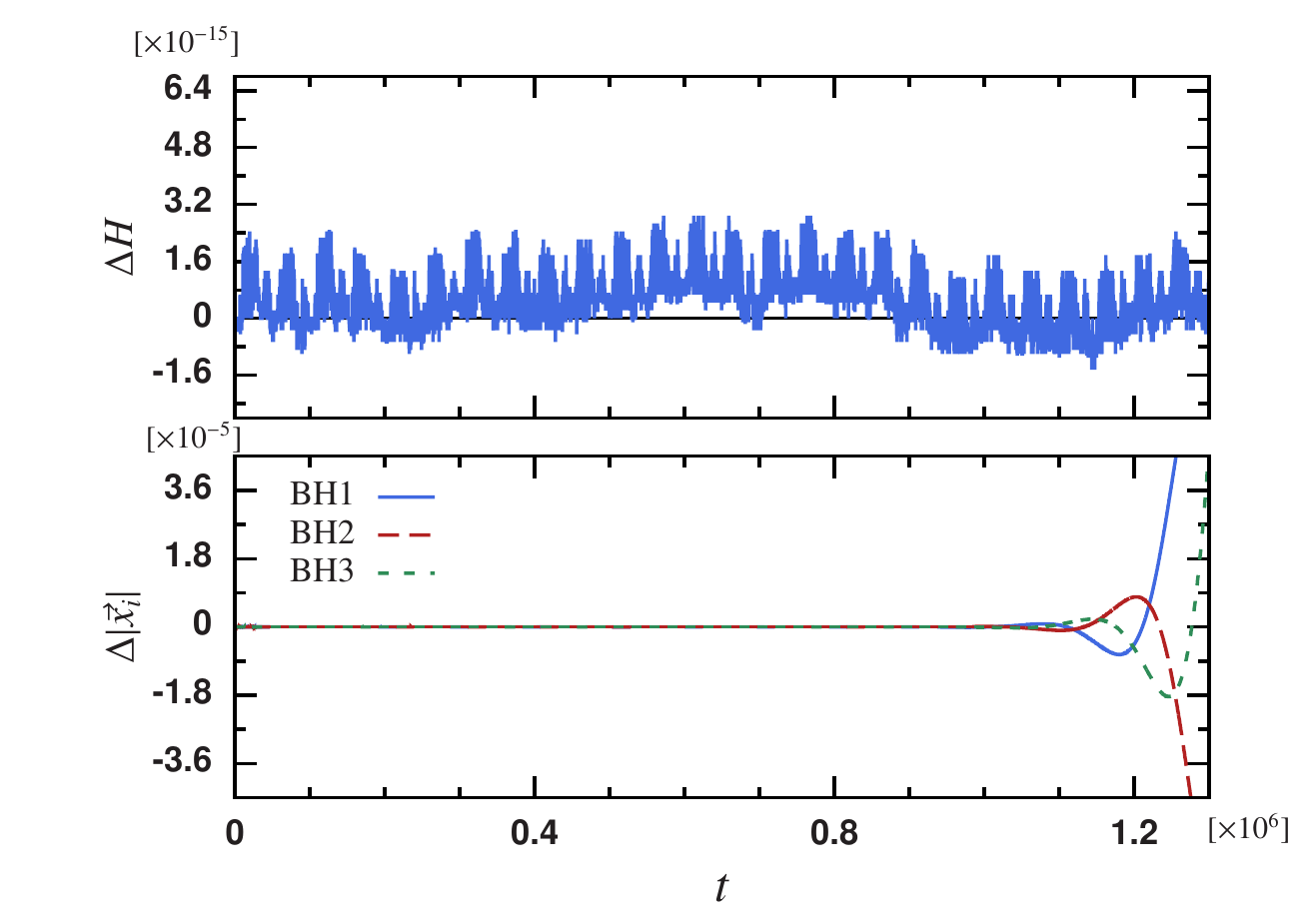}
  \caption{Test using Lagrange's equilateral solution of the Newtonian
    three-body  problem.   Shown  is  the relative  variation  of  the
    Hamiltonian (top) and the  relative change in the orbits (bottom).
  }
  \label{fig:1}
\end{figure}

\begin{figure}[btp]
  \centering
  \includegraphics[width=85mm]{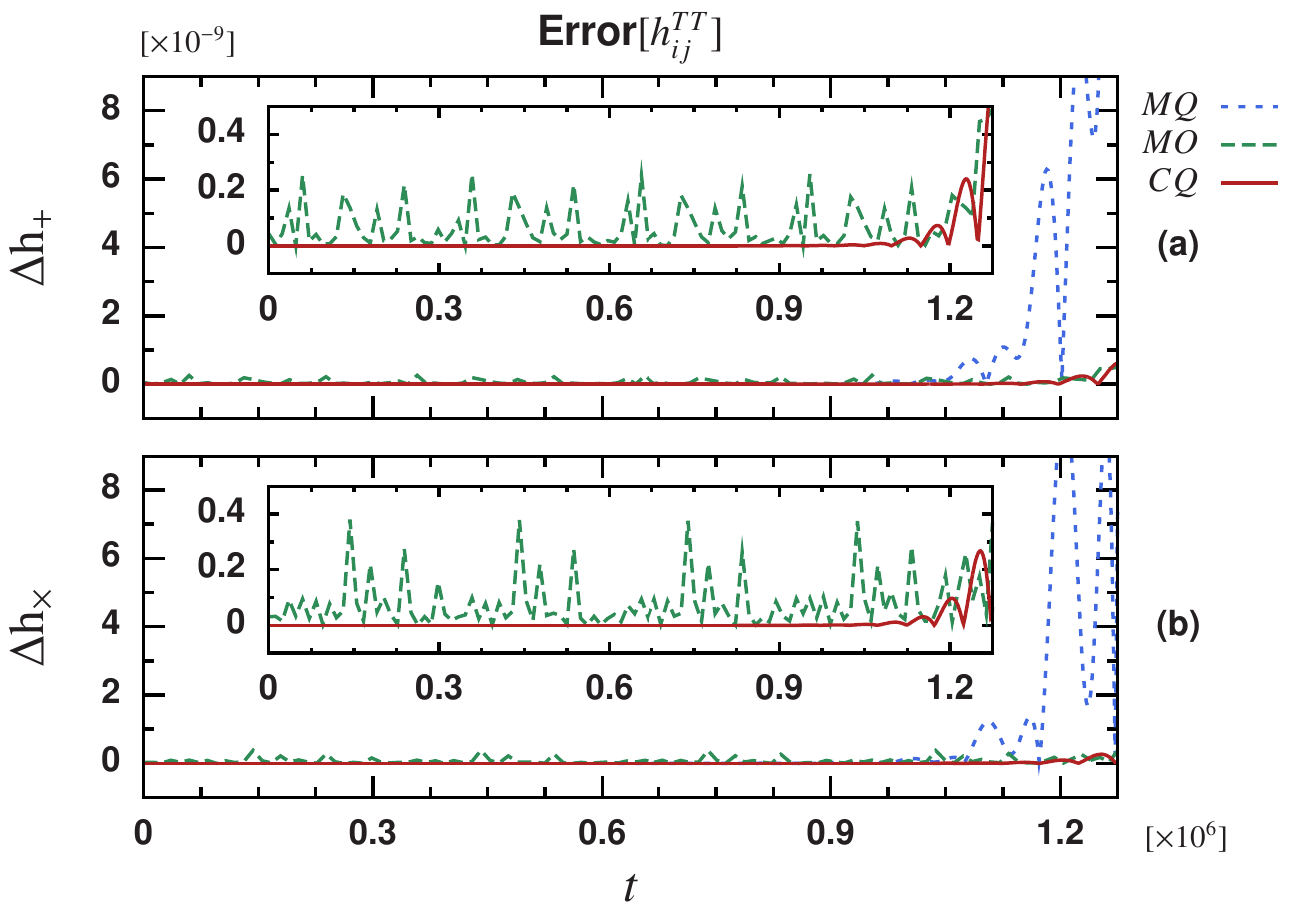}
  \caption{  Test   using  Lagrange's  equilateral   solution  of  the
    Newtonian three-body problem.  Shown  is the absolute value of the
    difference  between the  analytical expression  and  the numerical
    calculation for the mass quadrupole, mass octupole and the current
    quadrupole for each polarization of the waveform.  The insets show
    the mass octupole and the current quadrupole.  }
  \label{fig:2}
\end{figure}

In Fig.~\ref{fig:1}, we show the relative variation of the Hamiltonian
\begin{equation}\label{eq:33}
\Delta H := \frac{H(0)-H(t)}{H(0)}, 
\end{equation}
and for each body the  relative variation of the position with respect
to  the center of  mass.  The  variation of  the Hamiltonian  is small
(close to  machine accuracy),  however the error  in the  orbits grows
fast,  breaking the  regular  trajectory. In  this  case, after  seven
orbits  the numerical  solution fails.   The waves  exhibit  a similar
behavior.
In Fig.~\ref{fig:2},  we show the  error for each polarization  of the
waveforms  \eqref{eq:24}-\eqref{eq:26}.  The error  is defined  as the
absolute value of the difference between the numerical calculation and
the analytical  expression.  The mass octupole exhibits  a noisy error
due to  the complicated nature  of the analytical expression.   On the
other  hand, it seems  that the  error in  the mass  quadrupole starts
growing before the errors in the mass octupole and current quadrupole.
By looking at the analytical expressions this fact can be explained as
follows (see  Eqns.~(\ref{eq:59})-(\ref{eq:64})).  The mass quadrupole
part contains a factor $a^2  \omega^2$ (where $a$ is the separation of
the bodies and $\omega$ the orbital frequency).  The mass octupole and
current quadrupole have  a factor $a^3 \omega^3$. In  terms of $a$ the
factors  reduce to  $a^{-1}$  and $a^{-3/2}$  respectively.  A  small
change in the orbit is visible at a smaller length scale, and then the
growth in the waveforms seems to be delayed.

\begin{figure}[btp]
  \centering \includegraphics[width=85mm]{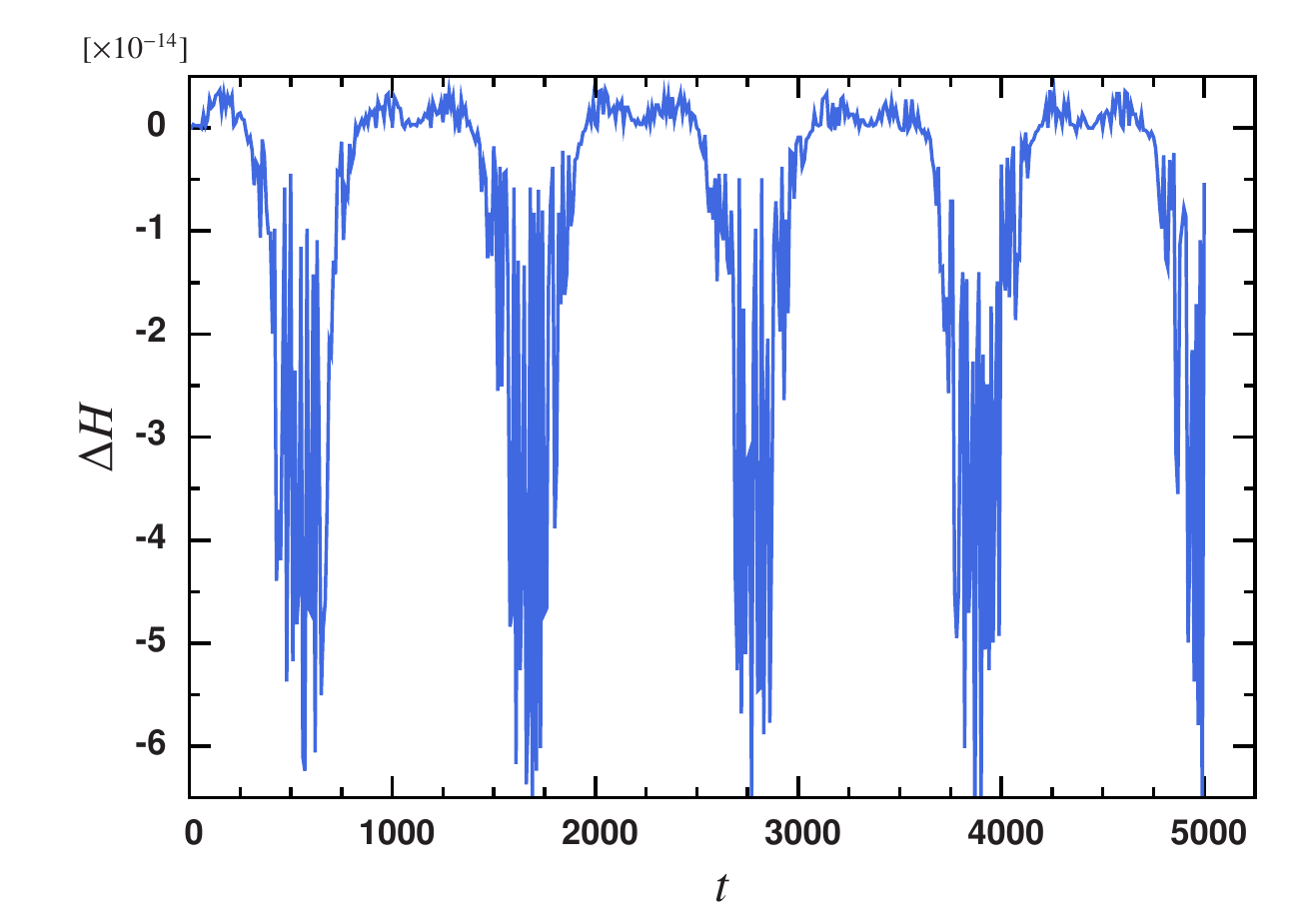} 
  \caption{Moore's figure  eight solution.  Relative  variation of the
    Hamiltonian    for     a    solution    which     includes    2~PN
    corrections.}  \label{fig:3}
\end{figure}


We reproduce  a few of the results  from \cite{LouNak08}, specifically
the simulation  of the  equal-mass Moore's figure  eight \cite{Moo93},
which  includes  first  and  second post-Newtonian  corrections.   Our
choice of method  was guided by numerical experiments  to minimize the
numerical  error  in this  example.   With  the double-step,  midpoint
method we  obtain fluctuations of  the Hamiltonian of  $10^{-14}$ (see
Fig.~\ref{fig:3}), while other methods and parameter settings can show
a significantly larger error.

\begin{figure}[btp]
  \centering
  \includegraphics[width=85mm]{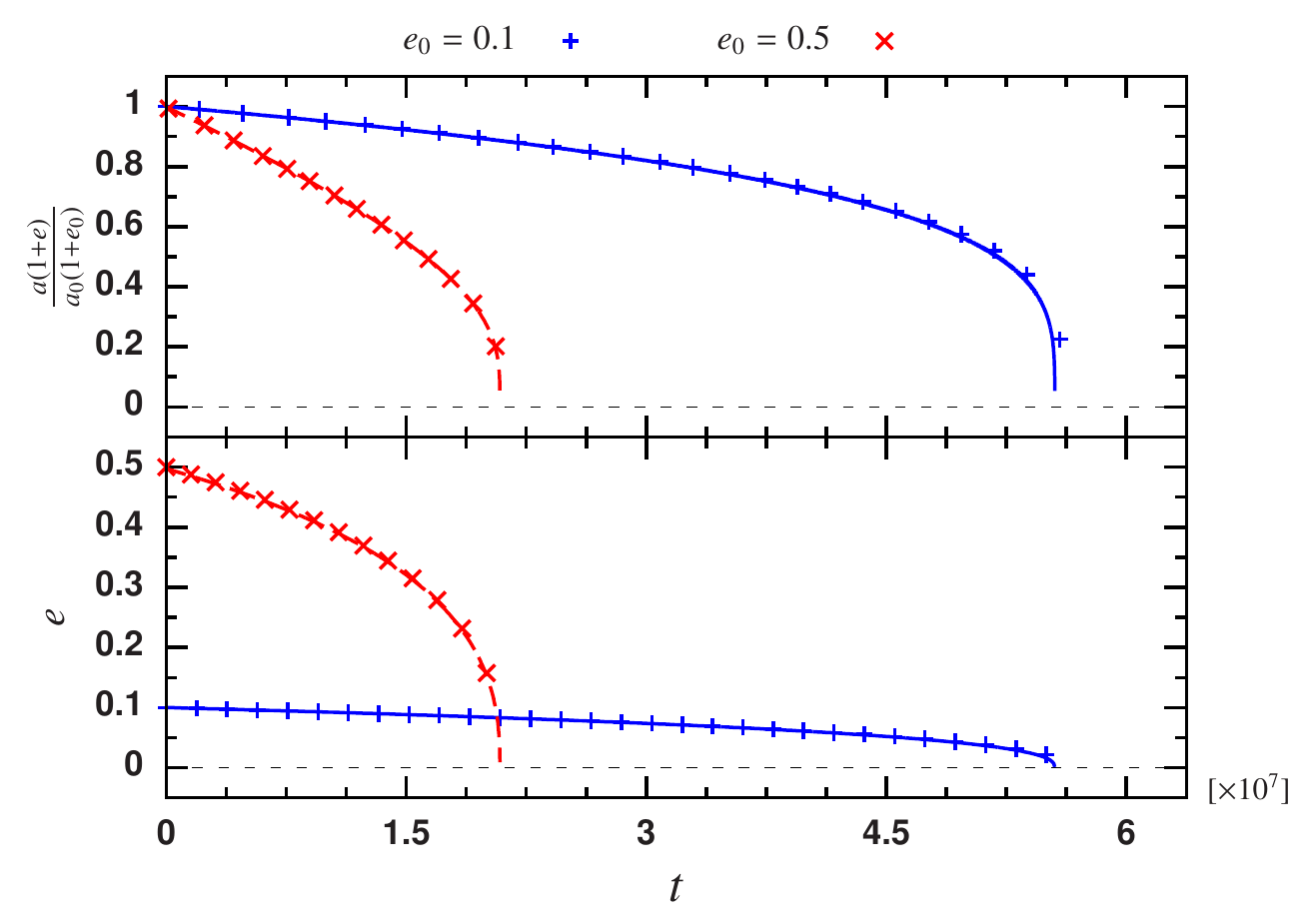}
  \caption{Binary  system  with   2.5~PN  radiation.   Top:  Relative
    variation of the apoapsis of the two bodies.  Bottom:
    eccentricity variation; comparison  of our numerical result (solid
    and   dashed   lines)    with   the   numerical   integration   of
    \eqref{eq:34} and \eqref{eq:35} (marks $+$ and $\times$)
    for two initial eccentricities.}
  \label{fig:5}
\end{figure}

We tested our $n-$body 2.5~PN  equations of motion for the case $n=2$,
i.e.\ for binary systems.  The  variation of the semimajor axis and of
the eccentricity of a binary system due to the gravitational radiation
is given by \cite{Pet64}
\begin{eqnarray}
\frac{\rd a}{\rd t} &=&
-\frac{64}{5}\frac{m_1m_2}{a^3(1-e^2)^{7/2}}\left(1+\frac{73}{24}e^2
+\frac{37}{96}e^4 \right),\label{eq:34}\\ 
\frac{\rd e}{\rd t} &=& -\frac{304}{15}\frac{m_1m_2}{a^4(1-e^2)^{5/2}}
\left(e +\frac{121}{304}e^3 \right).\label{eq:35}
\end{eqnarray}
We   tested  the   2.5~PN  equations   of  motion   \eqref{eq:12}  and
\eqref{eq:13} by  comparison with direct numerical  integration of the
Eqs.~\eqref{eq:34} and \eqref{eq:35}.
The test was  performed with two different binaries,  one with initial
eccentricity $e_0=0.1$ and  one with $e_0=0.5$.  In both  cases we set
$m_1=2  m_2$,  $a_0=160$.  The  numerical  integration  of the  2.5~PN
equations agree  very well with  the result provided by  the numerical
integration  of  \eqref{eq:34} and  \eqref{eq:35}.   We calculate  the
eccentricity of our orbits with the Newtonian formula
\begin{equation}\label{eq:36}
e=\sqrt{1+\frac{2 l^2 H_c}{(m_1 m_2)^3}},
\end{equation} 
where $l$ is the magnitude of  the total angular momentum and $H_c$ is
the value of  the conservative part of the  Hamiltonian.  The apoapsis
(the maximum separation of the two bodies) is related to the semimajor
axis by  $r_{\mathrm{ap}}=a(1+e)$.  For  simplicity we compare  in the
upper   panel   of   Fig.~\ref{fig:5}   the  relative   variation   of
$r_{\mathrm{ap}}$ to its initial value  and in the lower panel we show
the variation of the eccentricity.

\begin{figure}[btp]
  \centering 
  \includegraphics[width=85mm]{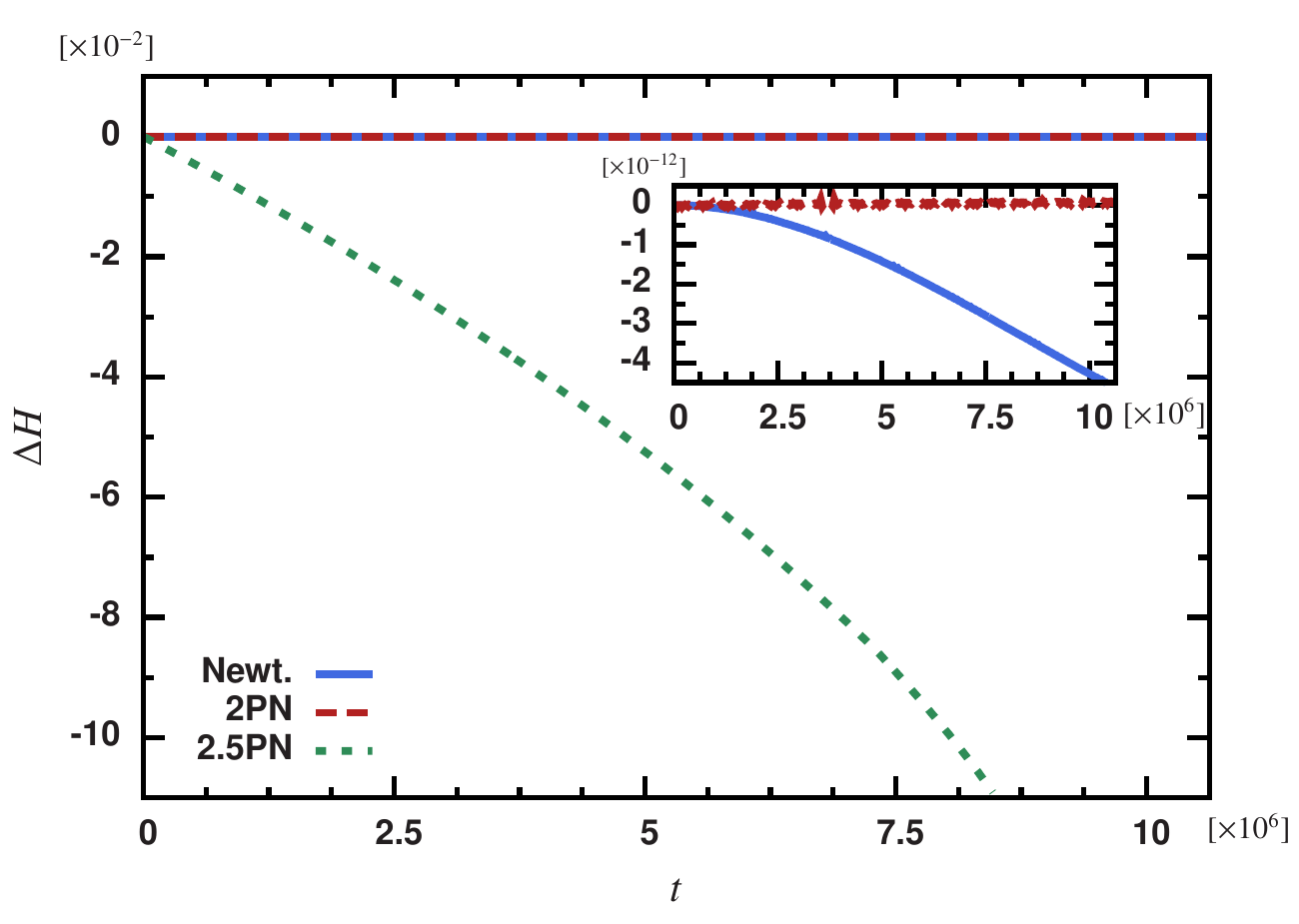}
  \caption{ H\'enon  criss-cross solution for Newtonian,  2 and 2.5~PN
    dynamics.   The main  panel shows  the relative  variation  of the
    Hamiltonian  for  the  three  cases.   The inset  shows  only  the
    conservative systems (Newtonian and 2~PN).  }
    \label{fig:6}
\end{figure}

\begin{figure}[btp]
  \centering
  \includegraphics[width=85mm]{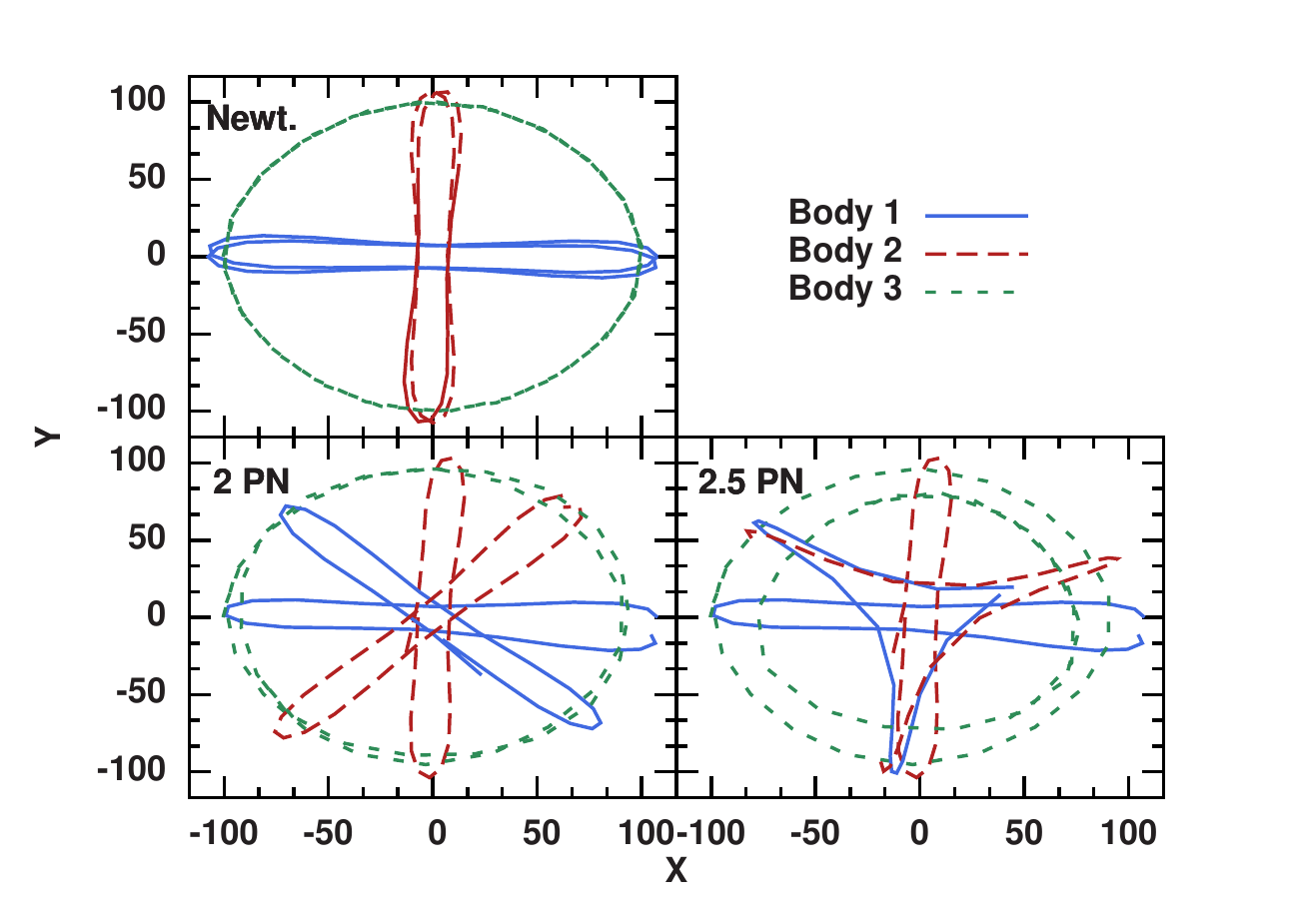}
  \caption{ H\'enon  criss-cross solution for Newtonian,  2 and 2.5~PN
    dynamics. First  and last orbits.   In the Newtonian  dynamics the
    orbits do  not show a significant change.   For dynamics including
    2~PN corrections the orbits  exhibit the expected precession.  For
    the dynamics  which includes 2.5~PN  corrections the gravitational
    radiation produces a significant change in the orbits which in the
    long run breaks the system.}
    \label{fig:7}
\end{figure}

In order to  test the code for long evolutions of  three bodies we use
H\'enon's   criss-cross   solution   \cite{Hen76,Moo93,Nau01}.    This
solution  is stable  with respect  to  a wide  range of  perturbations
\cite{MooNau08}.  We  evolve the  equal-mass criss-cross  solution for
around $10^3$ orbits  for ad-hoc initial parameters. In  our system of
units,
\begin{equation*}
\begin{array}{ll}
\vec{x}_1(0)=1.07590  \lambda^2 \hat{x}, & \vec{p}_1(0) = 3^{-3/2}
\cdot 0.19509 \lambda^{-1}\hat{y},\\  
\vec{x}_2(0)=-0.07095 \lambda^2 \hat{x}, & \vec{p}_2(0) = -3^{-3/2}
\cdot 1.23187 \lambda^{-1}\hat{y},\\
\vec{x}_3(0)=-1.00496 \lambda^2 \hat{x}, & \vec{p}_3(0) = 3^{-3/2} 
\cdot 1.03678 \lambda^{-1}\hat{y},
\end{array}
\end{equation*}
where $\hat{x}$, $\hat{y}$ and $\hat{z}$ are the unitary basis vectors
in Cartesian coordinates,  and $\lambda$ is a scaling  factor (for our
simulation  $\lambda=10$).  Notice  that  for  this test  we  use  the
parameters given in \cite{MooNau08} with the scaling factor $\lambda$,
and  doing a  change of  variables  from initial  velocity to  initial
momentum. Therefore,  we are not  including post-Newtonian corrections
to the  initial parameters.  In Fig.~\ref{fig:6} we  plot the relative
variation  of the  Hamiltonian  for the  evolution  using a  Newtonian
potential and  the corresponding Hamiltonian  variation for evolutions
which include 2 and 2.5~PN  corrections.  As is expected the variation
of  the  Hamiltonian  in the  2.5~PN  case  is  huge compared  to  the
conservative   case,   and    the   bodies   separate   after   around
$t=7.825\times10^6$.   The  inner panel  in  Fig.~\ref{fig:6} shows  a
detail of  the conservative  part.  In this  case the  2.5~PN dynamics
show  better  conservation  of  the  Hamiltonian in  contrast  to  the
Newtonian  case which  has a  variation in  the Hamiltonian  of around
$4\times 10^{-12}$.

We confirm that the system is stable even after the inclusion of 2 and
2.5~PN corrections,  see Fig.~\ref{fig:7}.  In the  Newtonian case the
accumulation  of numerical  errors  and probably  a  round-off in  the
initial parameters lead to a small variation of the orbits.  The basic
shape of  the criss-cross figure  suffers a small rotation.   The 2~PN
correction  includes  the effect  of  precession  in  the orbits;  the
original  figure spins  many times  around the  origin  preserving its
original  shape.  The  inclusion  of gravitational  radiation via  the
2.5~PN  corrections  has  a  stronger  effect on  the  orbits,  slowly
deforming  the original figure.   The body  in the  circularlike orbit
shows a  significant reduction  of the orbital  radius, the  two other
bodies follow at the end a triangular orbit with narrow corners.

We also use the Newtonian H\'enon criss-cross solution for performance
and accuracy tests. A  performance test based on walltime measurements
resulted in  about $4.4$ seconds per  orbit on (one core  of) an Intel
i7-860 processor. For accuracy testing,  we evaluate the error of time
integration by  a reversibility test.  After computing  a given number
of orbits, we solve the system backward in time starting with the last
position of each  particle but replacing every linear  momentum by its
opposite value.  To measure  the error we  compute the  differences in
phase space between the initial position and momentum and the position
and momentum  after the backward evolution. For  our standard setting,
the error after 100 orbits is on the order of $10^{-10}$.


\subsection{Strong perturbation of a binary system} 
\label{sec:strong-pert-binary}

Here we consider the strong  perturbation of the dynamics and waveform
of  a binary  compact object  system due  to a  third  smaller compact
object.  We take all PN corrections up to 2.5~PN for the three bodies.
This approach gives  us a good description of  the third body orbiting
close  to  the  binary.   However,  the  computational  cost  of  each
simulation increases with respect to the Newtonian simulations, making
it  too  costly  to  perform  a comprehensive  study  of  this  study.
Nevertheless, we  can select  a representative case  in an  attempt to
identify key properties.


\begin{figure}[btp]
  \centering
  \includegraphics[width=85mm]{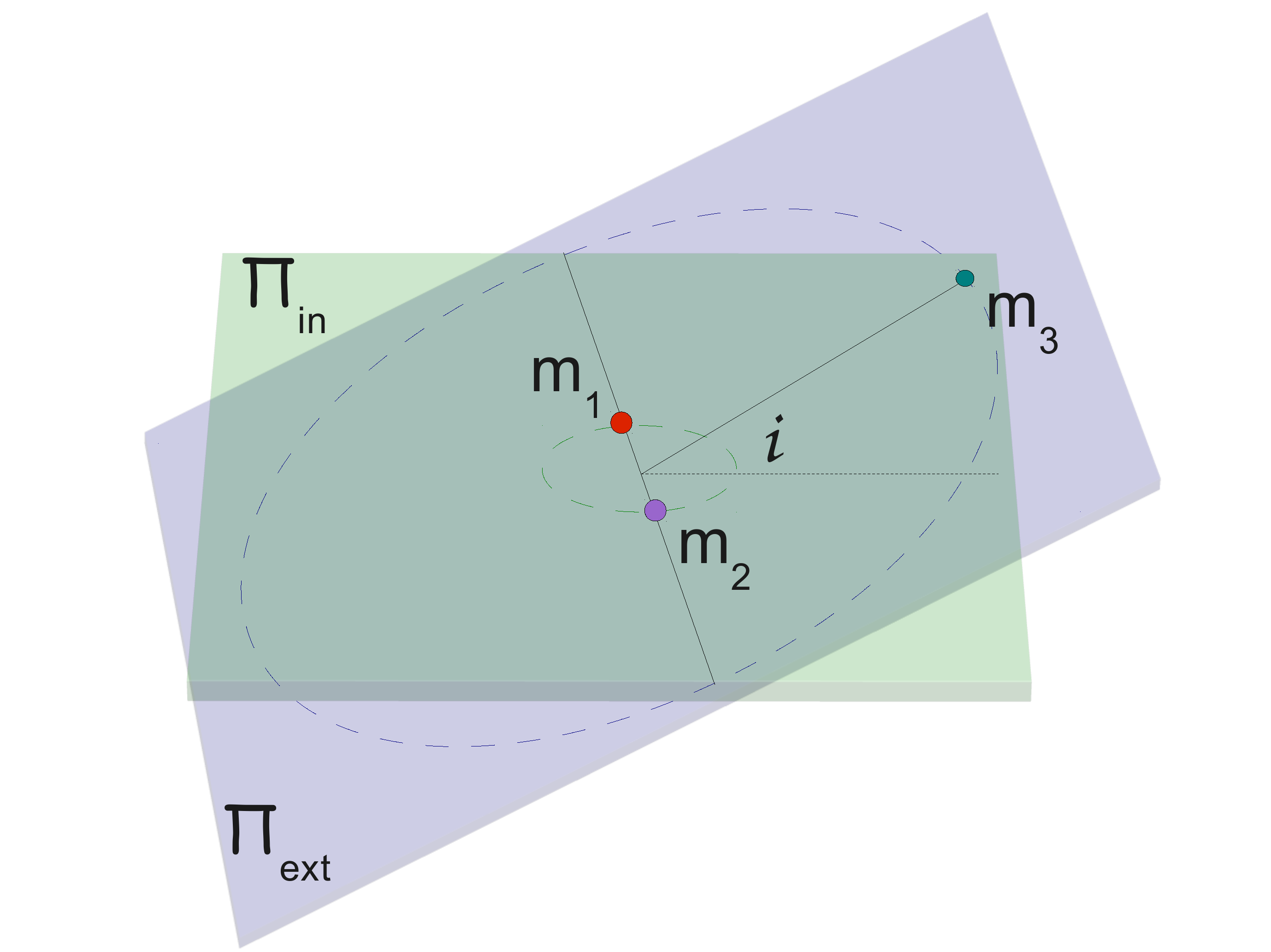}
  \caption{Hierarchical system. Initial configuration of the inner and
    external binaries. The initial momentum of the third body is given
    by considering  the external binary as a  Newtonian binary.  Shown
    are   the  osculating   orbital  planes   $\Pi_{\mathrm{in}}$  and
    $\Pi_{\mathrm{ext}}$  for inner and  external binary  orbits.  The
    two planes are inclined by an angle $i$.}
  \label{fig:9}
\end{figure}

As a  basic configuration we study  a Jacobian system  with mass ratio
10:20:1.  The inner binary  system has initial separation $r_b(0)=150$
and eccentricity $e_b(0)=0$. We set the initial parameters considering
only the Newtonian dynamics,  in particular, the eccentricity refers to
the Newtonian case.  We view the  third compact body and the center of
mass of the inner  binary as a new binary (we will  refer to it as the
external  binary).    The  external  binary   has  initial  separation
$r_3(0)=10000$ and initial  eccentricity $e_3(0)=0$.  The bodies start
from  a  configuration where  the  apoapsis  of  the inner  binary  is
perpendicular   to  the   apoapsis   of  the   external  binary   (see
Fig.~\ref{fig:9}).

\begin{figure}[btp]
  \centering
  \includegraphics[width=85mm]{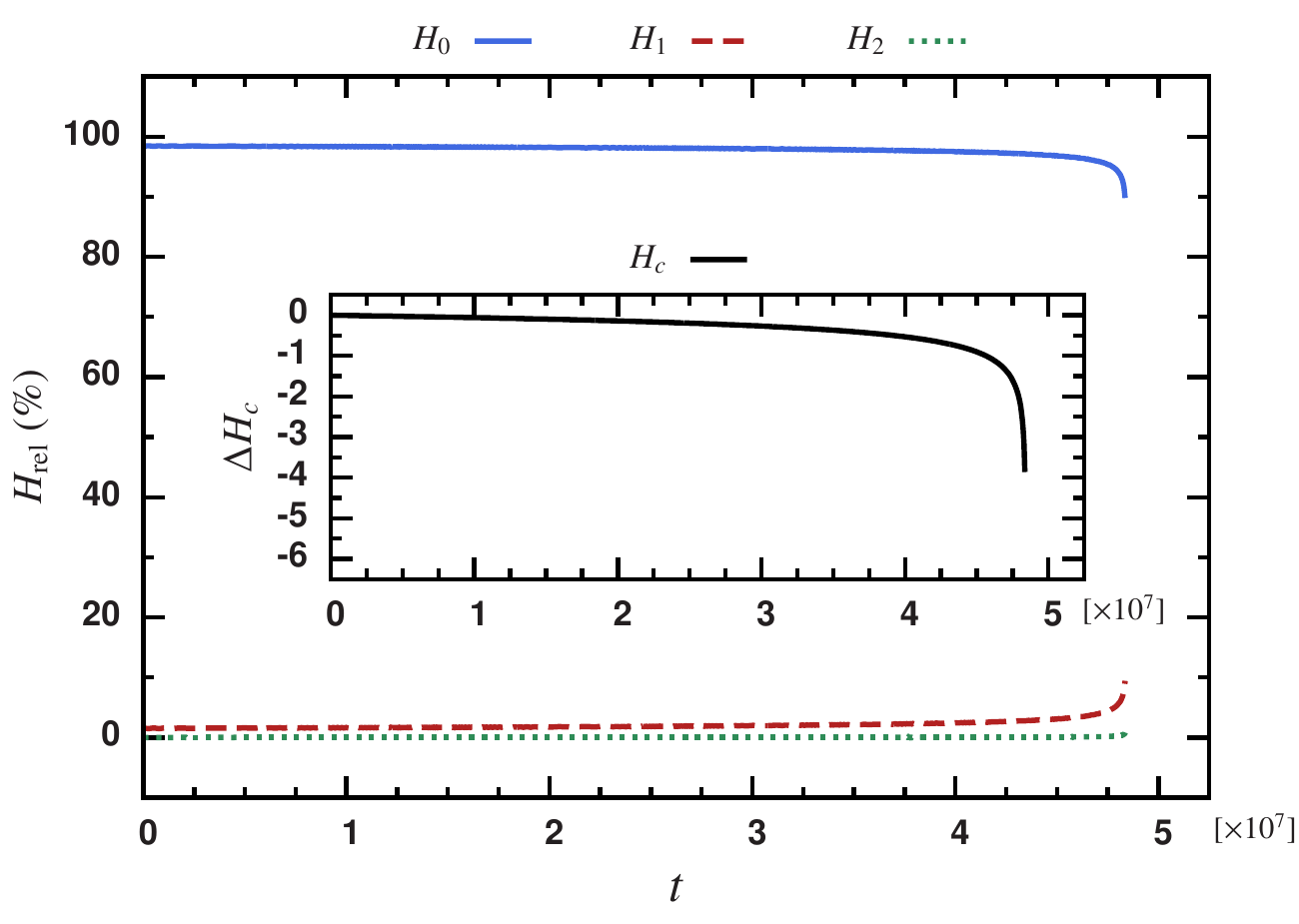}
  \caption{Planar hierarchical  system.  Relative contribution  to the
    Hamiltonian defined by \eqref{eq:32}.  The inset shows
    the time dependence  of  $H_c=H_0+H_1+H_2$. 
      Notice  that when  the  system
    approaches the merger phase, the Hamiltonian decreases quickly.  }
  \label{fig:10}
\end{figure}

We denote the inclination  angle between the osculating orbital planes
$\Pi_{\mathrm{in}}$    and    $\Pi_{\mathrm{ext}}$    by   $i$    (see
Fig.~\ref{fig:9}).   The behavior  of  the Hamiltonian  is similar  in
every case that we consider.  The conservative part of the Hamiltonian
decreases relatively  slowly during most of  the simulation.  However,
when the system approaches the merger phase, the Hamiltonian decreases
fast (see Fig.~\ref{fig:10}). As  we mentioned before, the simulations
are  stopped  when  the   contribution  of  the  first  post-Newtonian
correction becomes larger than 10\%. We consider this instant the time
when the merger phase starts.

We  consider  five  numerical  experiments.  In  Table~\ref{tab:1}  we
summarize the configurations of the numerical experiments. We vary one
parameter of the basic configuration and fix the rest.
The main goal of the study  is to characterize the changes produced in
the waveforms due to the change in each parameter.

\begin{table}[btp] 
  \begin{center}
    \caption{Configuration  of the  numerical experiments.   The fixed
      parameters  in  each case  are  the  mass  ratio $10:20:1$,  the
      initial eccentricity of the  inner binary $e_b=0$, and the angle
      between the apoapsis of the inner binary and the apoapsis of the
      external binary which is set to $\pi/2$.  The base configuration
      has    initial   binary   separation    $r_b=150$,   Hamiltonian
      $H_{0+1+2+2.5}$,  eccentricity of  the external  binary $e_3=0$,
      inclination  angle of  the osculating  planes $i=0$  and initial
      external binary separation $r_3=10000$}\label{tab:1}
    \begin{tabular*}{0.45\textwidth}{@{\extracolsep{\fill}}c|l}
      \hline 
      \hline 
      Experiment&
      Parameter variation
      \\ \hline
1 & $r_b\in\{130,140,150,160,170\}$ \\
      \hline 
2 & $H\in\{H_{0+2.5}, H_{0+1+2.5}, H_{0+1+2+2.5}\}$ \\
      \hline 
3 & $e_3 \in \{0,0.1,0.2,0.3,0.4,0.5,0.6\}$ \\
      \hline 
4 & $i \in \{0,\pi/8,\pi/4,3\pi/8,\pi/2\}$ \\
      \hline 
5 & $r_3 \in \{312.5, 625, 1250, 2500, 5000, 10000\}$ \\
      \hline 
      \hline
    \end{tabular*}
  \end{center}
\end{table}

\subsubsection{Binary versus triple system}
\label{sec:binary-triple-system}

\begin{figure}[btp]
  \centering
  \includegraphics[width=85mm]{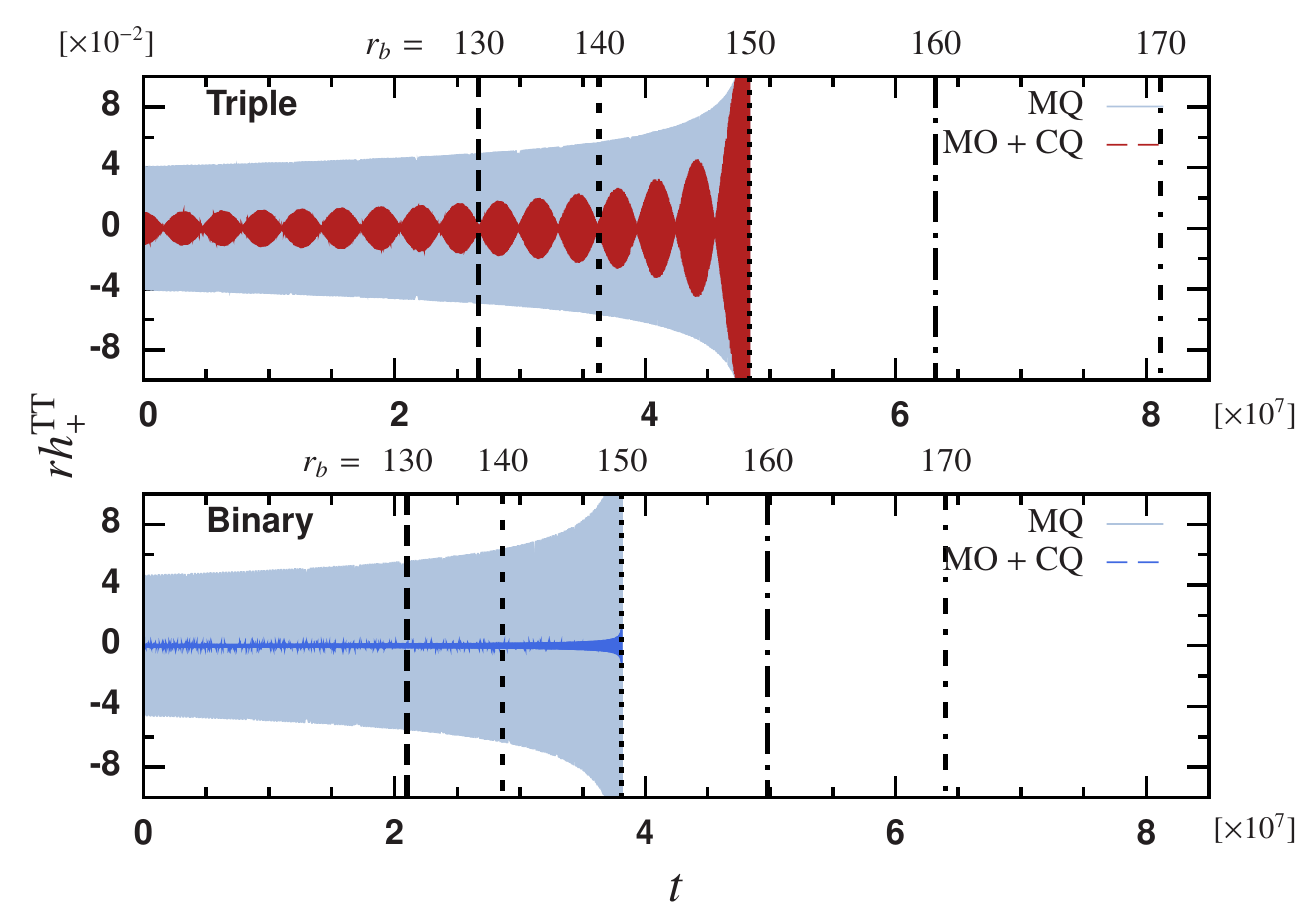}
  \caption{Planar   hierarchical  system.    Comparison   between  the
    perturbed binary  and the unperturbed one.   The light grey
    region is the mass quadrupole {MQ} contribution to the waveform $r
    h_{+}$, which  is not resolved  since there are 4070  orbits.  The
    dark region  inside the light  grey one is  the mass
    octupole plus  the current quadrupole {MO+CQ}  contribution to the
    waveform  $r h_{+}$.  The vertical  lines mark  the time  when the
    simulations     are     stopped     for     initial     separation
    $r_3\in\{130,140,150,160,170\}$ of the inner binary.}
  \label{fig:11}
\end{figure}

We compare the  case where the inner binary is  not being perturbed by
the third  compact body.  Fig.~\ref{fig:11} shows  the components of
the  waveform  for  the   $h_+$  polarization  with  an  observational
direction  $\theta=\pi/4, \phi=0$.  In  both cases  the plot  shows in
light grey the mass quadrupole.
The  waveform looks  like  a  shadow region  because  compared to  the
timescale of  the entire  evolution a single  cycle looks like  a very
high frequency wave.
In Fig.~\ref{fig:11}, the binary  has completed 3448 orbits, while the
inner binary  of the triple system  has completed 4071  orbits and the
outer binary has completed 7.5 orbits.
The  mass octupole  plus the  current  quadrupole MO+CQ  are the  dark
region.  Notice  that in the triple  system MO+CQ is  modulated by the
period of the third body  (one cycle of modulation corresponds to half
an orbit of the third body).  The perturbation furthermore affects the
merger time;  for the triple system  it takes more time  for the inner
binary to  merge.  We  run the simulation  for 5 initial  inner binary
separation  $r_b\in\{130,140,150,160,170\}$.  In  Fig.~\ref{fig:11} we
mark  with  vertical lines  the  time  at  which the  simulations  are
stopped.  The relative change in the merger time
\begin{equation}\label{eq:37}
\frac{t_{\mathrm{3BH}}-t_{\mathrm{2BH}}}{t_{\mathrm{2BH}}}=0.270\pm0.0025,
\end{equation}
is  almost constant for  this simulations  (the standard  deviation is
0.0025).   We  did  not  observe  any particular  differences  in  the
waveform when changing $r_b$.

\subsubsection{Post-Newtonian corrections}
\label{sec:post-newt-corr}

\begin{figure}[btp]
  \centering
  \includegraphics[width=85mm]{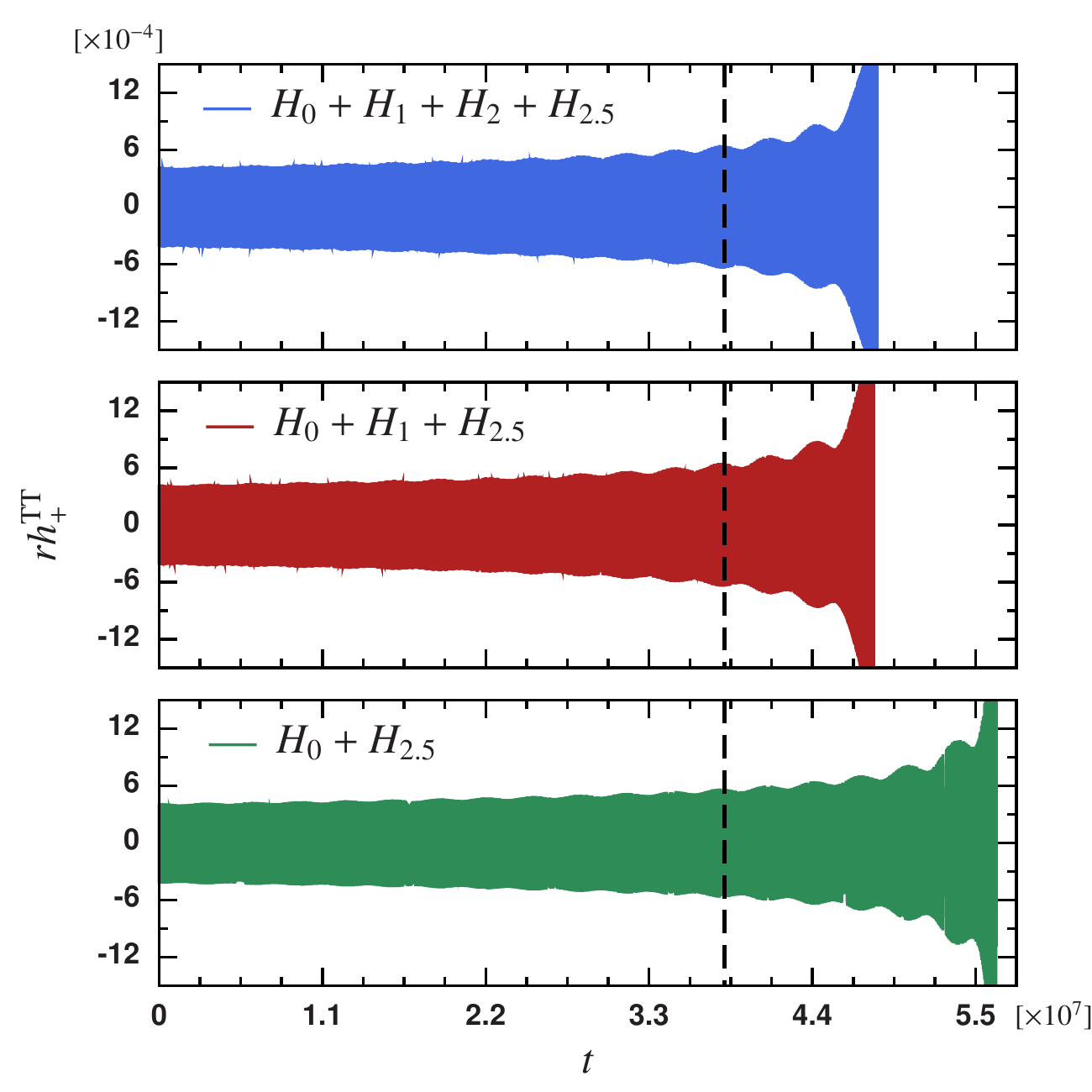}
  \caption{Successive  changes in the  waveform due  to post-Newtonian
    corrections.  Waveform  of a radiative  Newtonian system (bottom),
    radiative 1~PN system (middle), and full 2.5~PN system (top).  The
    waveform  includes the current  and mass  quadrupole and  the mass
    octupole contributions.   The vertical dash  line at $t=3.81\times
    10^7$ mark  the time when  the non-perturbed binary  system enters
    the merger phase (see Fig.~\ref{fig:11}).}
  \label{fig:12}
\end{figure}

In addition  to the comparison  to the nonperturbed binary  system, we
use  the  planar  configuration   to  explore  the  influence  of  the
conservative post-Newtonian corrections.  As in the previous case with
initial  binary   separation  $r_b=150$  (which  we   will  denote  as
\textit{full  2.5~PN} case),  we  solve the  system  for equations  of
motion   where  we   remove   the  2~PN   part   of  the   Hamiltonian
(\textit{radiative  1~PN})  and  where  we  remove  both  1  and  2~PN
corrections (\textit{radiative Newtonian}).  The full 2.5~PN case does
not show  a big difference compared  to the radiative  1~PN case.  The
merger phase time changes from $t=4.8372\times 10^7$ in the first case
to $t=4.8132\times  10^7$ in  the second one.   The waveform  does not
suffer  a noticeable  change  (see Fig.~\ref{fig:12}).   On the  other
hand,   in   the  radiative   Newtonian   case   the  result   changes
significantly.  The  merger phase time  starts later than  in previous
cases (around $t=5.6388\times 10^7$).
For  this  configuration  dynamic  which include  the  radiative  1~PN
corrections seems to be a good approximation. However, for the rest of
the simulations we employ the full 2.5~PN corrections.

\subsubsection{Variation of the eccentricity of the external binary}
\label{sec:vari-eccentr-extern}
\begin{figure}[btp]
  \centering 
  \includegraphics[width=85mm]{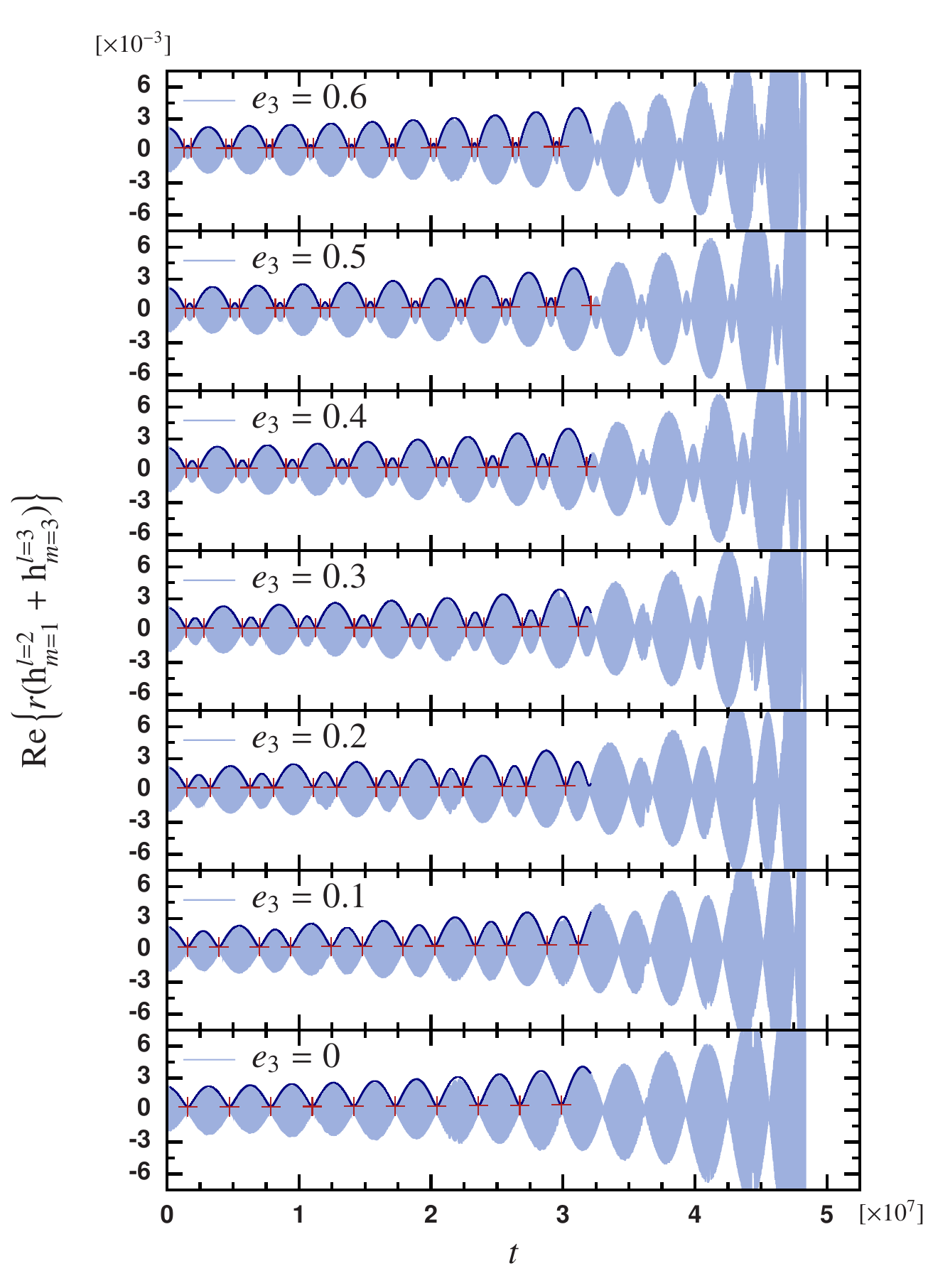}
  \caption{Sum  of modes  $l=2,m=1$  and $l=m=3$  as  function of  the
    eccentricity $e_3$.  From bottom to top the variation of the modes
    for $\e_e\in\{0,0.1,\dots,0.6\}$. The dark line is the envelope of
    the function  for 2/3 of the  total simulation. The  $+$ marks are
    the local minima of the envelope.}
  \label{fig:13}
\end{figure}

We  analyzed  the variation  of  the waveform  as  a  function of  the
eccentricity  of the external  binary $e_3$.   We ran  simulations for
$\e_3\in\{0,0.1,\dots,0.6\}$.   In  this  case  the  response  to  the
variation of  the eccentricity is better reflected  in the combination
of                     $\mathrm{h}^{l=2}_{m=1}$                    and
$\mathrm{h}^{l=3}_{m=3}$.   Figure~\ref{fig:13}  shows   the   sum  of
$l=2,m=1$  and $l=m=3$  modes  (which are  the  leading components  of
current quadrupole  and mass octupole,  respectively).  The modulation
of the  modes shows two  characteristic low-frequencies. If  we divide
the orbit of the external binary in two parts, one defined by the true
anomaly\footnote{The  true  anomaly  is  defined as  the  angle  which
  connects the  periapsis, the  main focus and  the trajectory  of the
  reduced   body.}   $\varphi$   running   from  $\varphi=-\pi/2$   to
$\varphi=\pi/2$     and     the     other    by     the     complement
$\varphi\in[\pi/2,3\pi/2]$,   it   is   possible  to   associate   the
characteristic frequencies to each  part of the trajectory. We compute
the  envelope of the  absolute value  of the  signal using  a low-pass
filter (dark  line in Figure~\ref{fig:13})  for 2/3 of the  total signal
(that part of the signal was easier to process for high eccentricity).
Using the resulting function  we compute numerically the local minima.
The differences between minima  are associated with the characteristic
frequencies.   An  alternative   way  to  extract  the  characteristic
frequencies  is by  looking at  the  Fourier spectra  of the  filtered
waveform.

We  label   the  period  for   $\varphi\in[-\pi/2,\pi/2]$  as  $\Delta
t_{\mathrm{ap}}$  and  the  period for  $\varphi\in[\pi/2,3\pi/2]$  as
$\Delta t_{\mathrm{per}}$ (at  $\varphi=0$ the external binary reaches
the  periapsis and at  $\varphi=\pi$ the  apoapsis). Table~\ref{tab:2}
shows the results, where we include the quotient.

\begin{table}[btp] 
  \begin{center}
    \caption{Periods    $\Delta    t_{\mathrm{per}}$    and    $\Delta
      t_{\mathrm{ap}}$ and its quotient. The values are computed using
      the  averages  of  the   differences  between  the  minima  (see
      Figure~\ref{fig:13})    and   the    errors    by   the    standard
      deviation.}\label{tab:2}
    \begin{tabular*}{0.45\textwidth}{@{\extracolsep{\fill}}l|ccc}
      \hline \hline   $e_3$    &    $\Delta    t_{\mathrm{per}}\;[\times10^6]$&    $\Delta
      t_{\mathrm{ap}}\;[\times10^6]$      &      $\Delta     t_{\mathrm{per}}/\Delta
      t_{\mathrm{ap}}$    \\  \hline
      0  &	$3.1473\pm0.00020$&	$3.1427\pm0.00053$&	$0.9985\pm0.00023$\\
      0.1&	$2.3812\pm0.00092$&	$3.0700\pm0.00120$&	$1.2890\pm0.00100$\\
      0.2&	$1.7890\pm0.00180$&	$2.9950\pm0.00051$&	$1.6750\pm0.00190$\\
      0.3&	$1.3260\pm0.00170$&	$2.9160\pm0.00110$&	$2.2000\pm0.00360$\\
      0.4&	$0.9590\pm0.00160$&	$2.8370\pm0.00110$&	$2.9580\pm0.00610$\\
      0.5&	$0.6690\pm0.00110$&	$2.7530\pm0.00170$&	$4.1180\pm0.00920$\\
      0.6&	$0.4390\pm0.00330$&	$2.6670\pm0.00400$&	$6.0700\pm0.05500$\\
      \hline 
      \hline
    \end{tabular*}
  \end{center}
\end{table}

In   the  Newtonian   case   it  is   possible   to  compute   $\Delta
t_{\mathrm{ap}}$ and $\Delta  t_{\mathrm{per}}$ using the conservation
of  the  angular momentum  $l$  and the  equation  of  the orbit  (see
e.g.\ \cite{GolPooSaf01}). The result is
\begin{eqnarray}\label{eq:38}
\Delta t_{\mathrm{per}}&=&\frac{l^3}{\mu}\int_{\pi/2}^{3\pi/2}(1+e \cos \varphi)^{-2}\rd \varphi,\\ 
\Delta t_{\mathrm{ap}}&=&\frac{l^3}{\mu}\int_{-\pi/2}^{\pi/2}(1+e \cos \varphi)^{-2}\rd \varphi, 
\end{eqnarray}
where $\mu$ is the reduced mass  of the binary.  The quotient between the
periods is related to the eccentricity by
\begin{equation}\label{eq:39}
\frac{\Delta t_{\mathrm{per}}}{\Delta t_{\mathrm{ap}}} = \frac{\pi}{2 \arctan \sqrt{\frac{1-e}{1+e}}-e \sqrt{1-e^2}}-1.
\end{equation}

\begin{figure}[btp]
  \centering 
  \includegraphics[width=85mm]{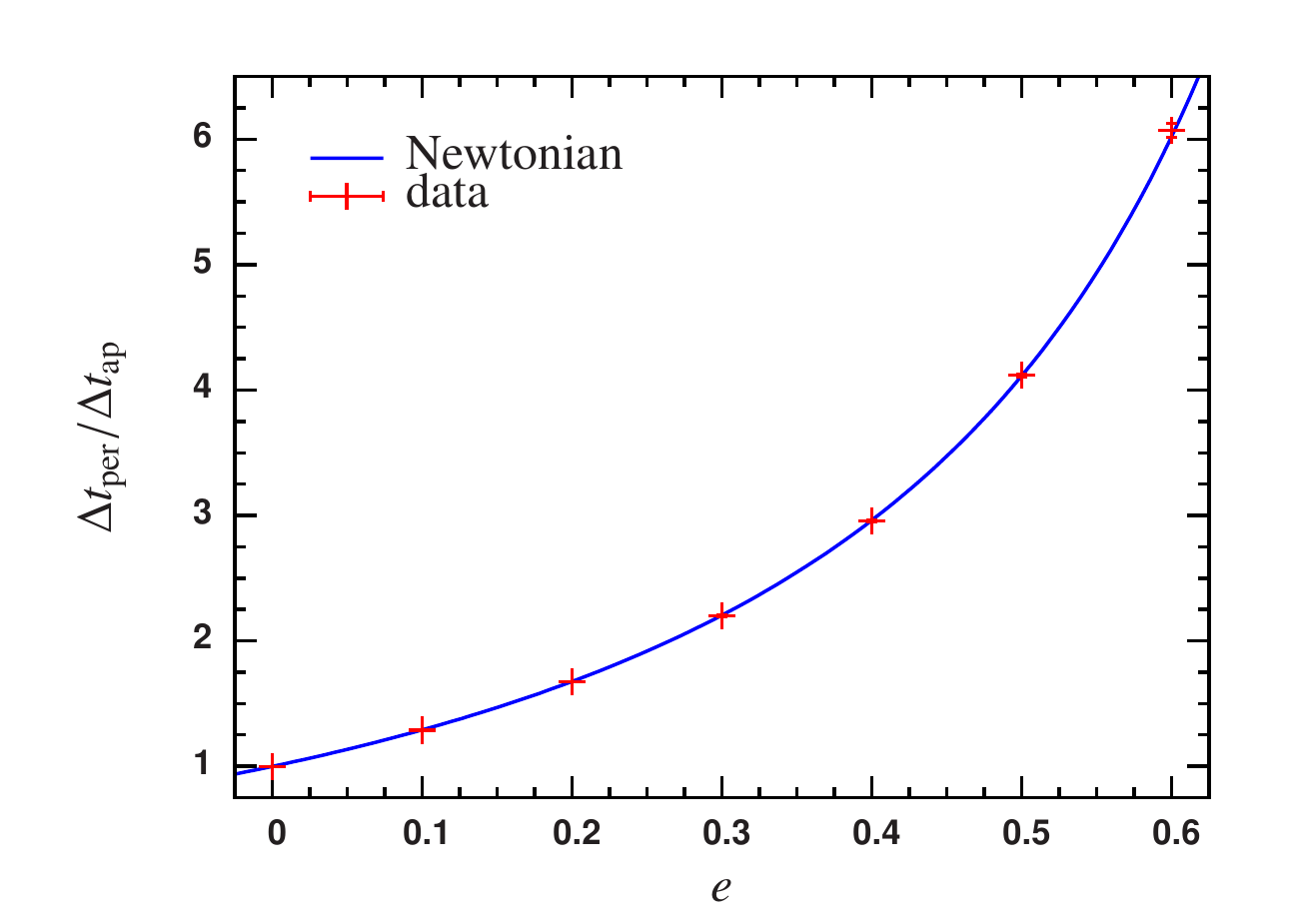}
  \caption{Ratio of the  periods $\Delta t_{\mathrm{per}}$ and $\Delta
    t_{\mathrm{ap}}$ as function of  the eccentricity.  The solid line
    shows the Newtonian relation.}
  \label{fig:14}
\end{figure}

Figure~\ref{fig:14} shows  a comparison between the  data presented in
Table~\ref{tab:2}  and the  Newtonian  expression \eqref{eq:39}.   For
this case the Newtonian expression represents very well the functional
behavior of our simulation.

\subsubsection{Variation of the inclination angle}
\label{sec:vari-incl-angle}

\begin{figure}[btp]
  \centering
    \includegraphics[width=85mm]{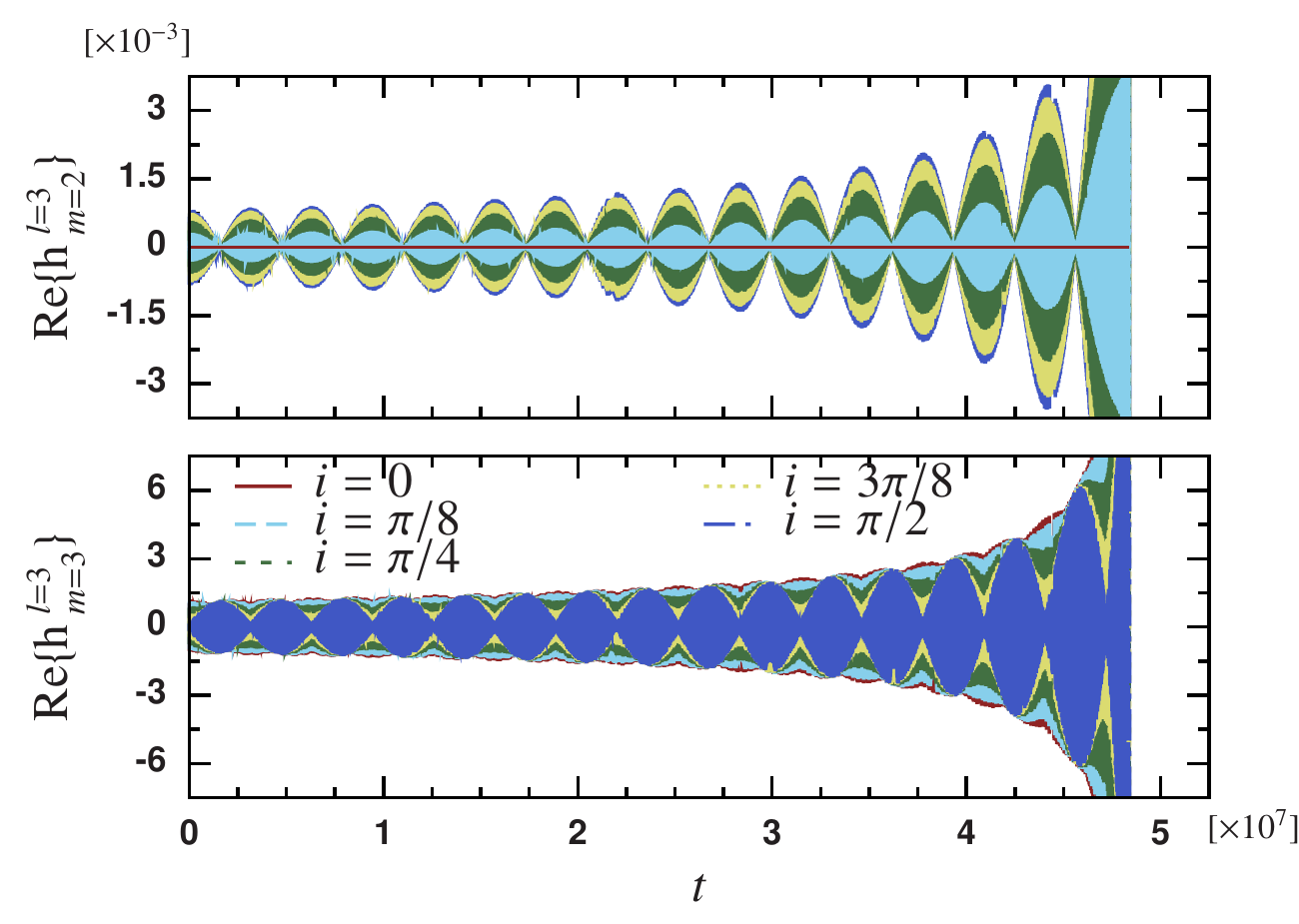}
  \caption{Variation  of the  amplitude  of $l=3$,  $m=2,3$, modes  as
    a function  of the  inclination angle  $i$.  
    Superposition   of  $\mathrm{Re}\{   \h^{\;l=3}_{\;m=2}   \}$  and
    $\mathrm{Re}\{   \h^{\;l=3}_{\;m=3}  \}$   as  function   of  $i$.}
  \label{fig:15}
\end{figure}

The  period of  modulation  of the  $l=3$  modes of  the waveform  are
related  to the  period of  the third  body.  On  the other  hand, the
amplitude  of the $l=3$  spherical components  of the  waveform encode
information about  the inclination angle $i$. We  run simulations with
the  same initial  configuration for  $i\in\{0, \pi/8,  \pi/4, 3\pi/8,
\pi/2\}$.   Figure~\ref{fig:15} shows the  variation of  the amplitude
for   the   real   part   of  the   modes   $\h^{\;l=3}_{\;m=2}$   and
$\h^{\;l=3}_{\;m=3}$ as  a function  of $i$.  Since  the real  and the
imaginary part of the modes  show the same behavior, for simplicity we
present  only  the  analysis of  the  real  part.   The real  part  of
$\h^{\;l=3}_{\;m=2}$ is  zero for  planar motion $i=0$.   However, the
contribution of this mode increases  with $i$.  On the other hand, the
contribution of $\mathrm{Re}\{\h^{\;l=3}_{\;m=3}\}$  is maximal in the
planar  case  and decreases  when  $i$  increases.   This behavior  is
symmetric with respect to $i=\pi/2$ and periodic with period $\pi$.

\begin{figure}[btp]
  \centering
  \includegraphics[width=85mm]{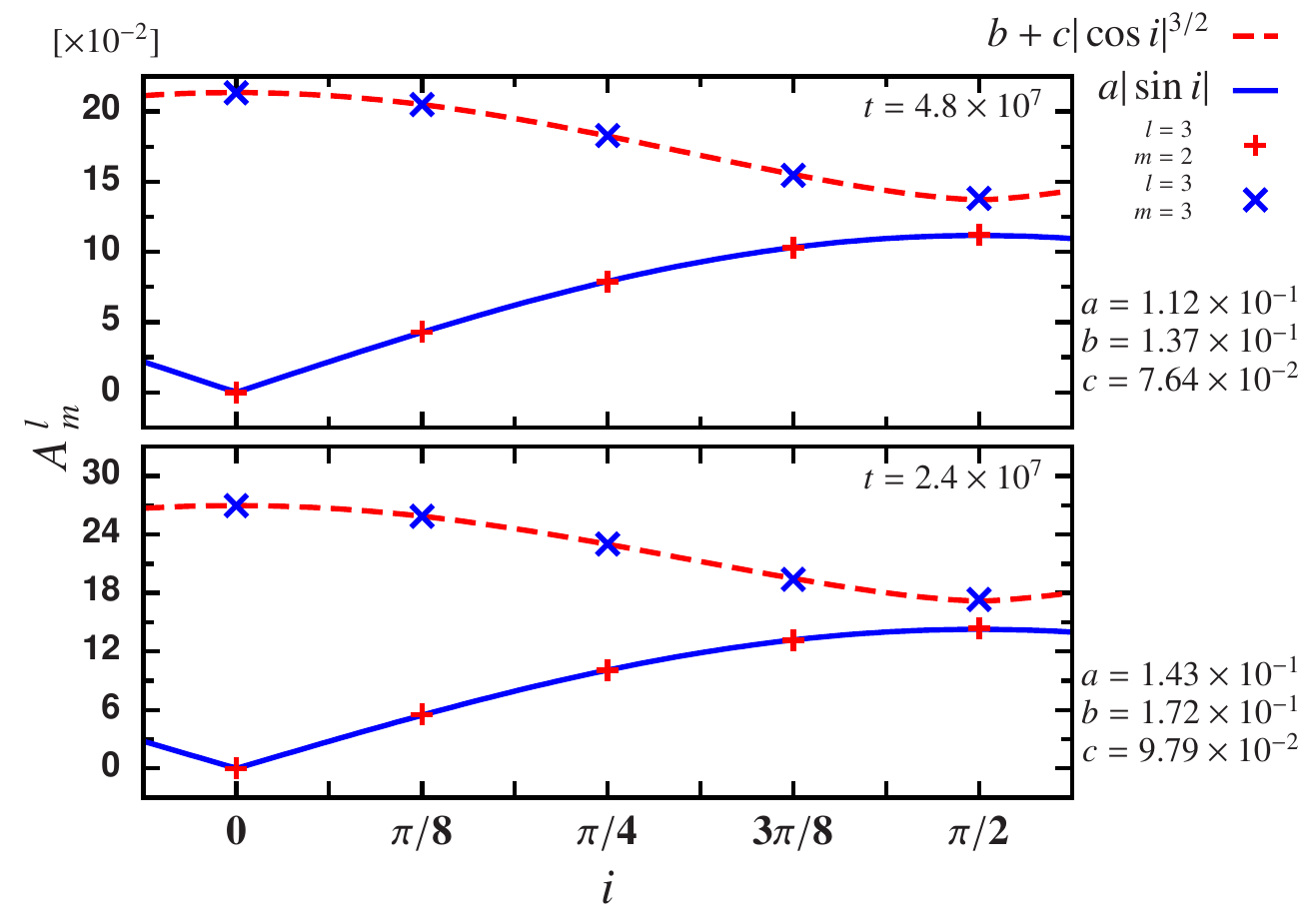}
  \caption{  Variation  of  $A^{\,l}_{\,m}$  as a function  of  $i$  for
    $t=t_f$ (upper panel) and $t=t_f/2$ (lower panel). }
  \label{fig:16}
\end{figure}

We   estimate  the   contribution   of  each   mode  calculating   the
\textit{area} which is covered by the real part of the mode,
\begin{equation}\label{eq:40}
\mathcal{A}^{\,l}_{\,m}(\tau):=   -   \int_{t_f}^{\tau}   \vert
\mathrm{Re}\{\h^{\,l}_{\,m}(\bar{\tau})\} \vert \, \rd \bar{\tau},
\end{equation}
where $t_f=4.8372 \times 10^7$ is  the final time of the evolution and
$\tau=t_f-t$.   We  integrate  backward  in  time  starting  with  the
beginning of the merger phase at $t_f$.
We compute  $\mathcal{A}^{\,l}_{\,m}(t)$ for 8  uniformly spaced times
during  the simulation.  We  normalize the  results using  the maximum
value   $\mathcal{A}_{\mathrm{max}}=\mathcal{A}^{\,l=2}_{\,m=2}$.   We
denote the normalized area by  $A^{\,l}_{\,m}$.  As an example we show
the  results for $\tau=0$  in Table~\ref{tab:3}  where we  present the
relevant modes.  In total we  compute 8 tables similar to the previous
one, however for brevity we do not present them here.  Notice that the
contribution of the $l=2$ modes is almost constant with respect to the
inclination angle $i$.  In  Figure~\ref{fig:16} we show the variation of
$A^{\,l=3}_{\,m=2}$ and $A^{\,l=3}_{\,m=3}$ for two integration times,
$\tau=0$ and $\tau = t_f/2$.

\begin{table}[btp] 
  \begin{center}
    \caption{Variation  of  $A^{\,l}_{\,m}$   as  a  function  of  the
      inclination angle $i$.  }\label{tab:3}
    \begin{tabular*}{0.45\textwidth}{@{\extracolsep{\fill}}cc|lllll}
      \hline   \multicolumn{2}{r|}{$\tau=0$}   &   $i=0$&
      $i=\pi/8$ & $i=\pi/4$ &  $i=2\pi/8$ & $i=\pi/2$ \\ \hline \hline
       $l$ & $m$ & \multicolumn{5}{c}{ $A^{\,l}_{\,m}$}\\ \hline
2&	0&	0.0019&	0.0019&	0.0021&	0.0024&	0.0026\\
2&	1&	0.0000&	0.0007&	0.0013&	0.0016&	0.0018\\
2&	2&	1.0000&	1.0000&	1.0000&	1.0000&	1.0000\\
3&	0&	0.0000&	0.0006&	0.0021&	0.0012&	0.0544\\
3&	1&	0.0546&	0.0527&	0.0588&	0.0397&	0.1160\\
3&	2&	0.0000&	0.0429&	0.0799&	0.1033&	0.1583\\
3&	3&	0.2128&	0.2052&	0.1957&	0.1552&	0.2376\\
      \hline 
      \hline
    \end{tabular*}
  \end{center}
\end{table}

\begin{figure}[btp]
  \centering 
  \includegraphics[width=85mm]{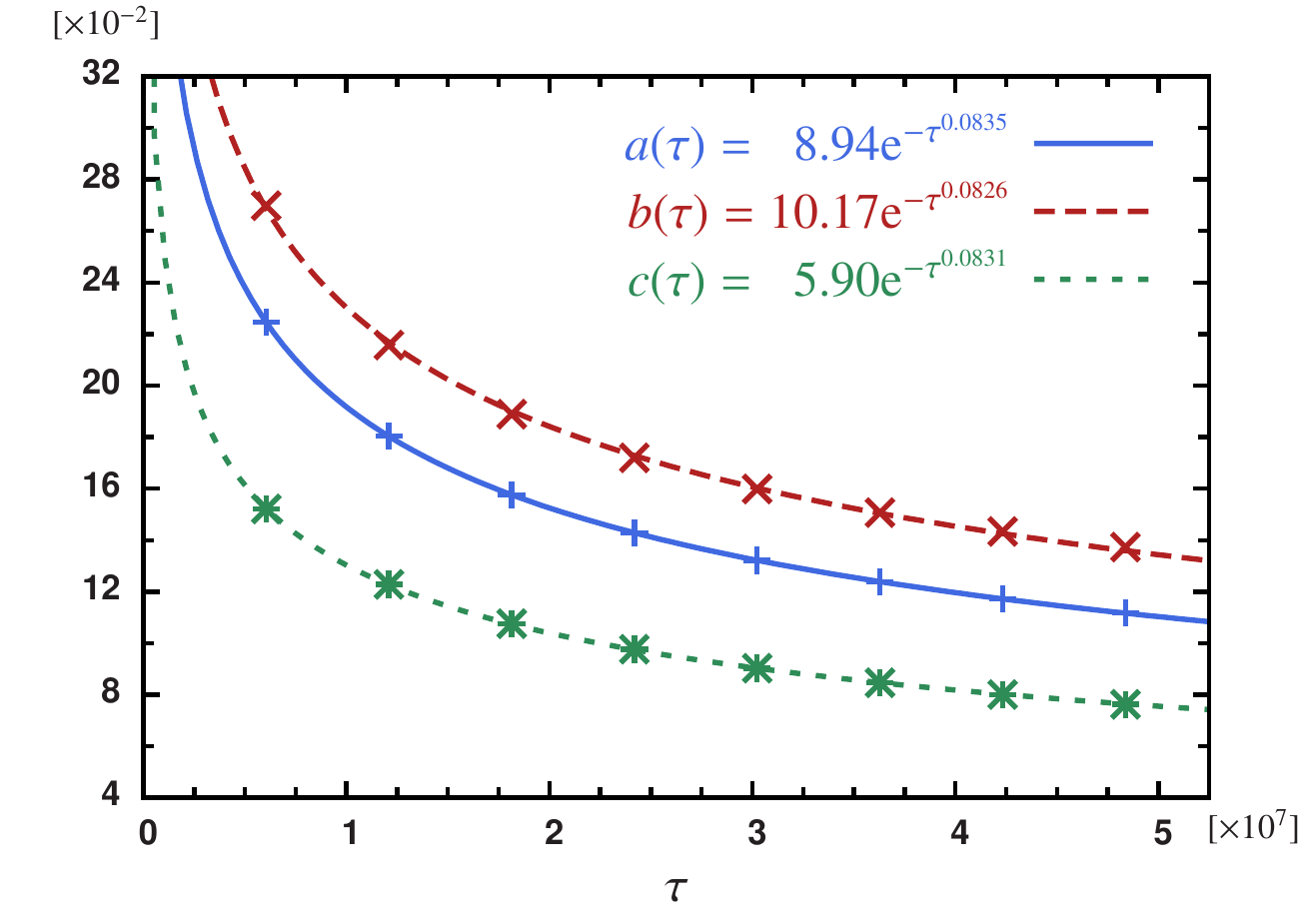}
  \caption{Functional  behavior  of  the fitting  coefficients.   The
    coefficients are  well described by an  exponential decay function
    in $\tau$.  }
  \label{fig:17}
\end{figure}

We found that the variation of $A^{\,l=3}_{\,m=2}$ is well represented
by
\begin{equation}\label{eq:41}
A^{l=3}_{m=2}(t,i)=a(\tau) \vert \sin i \vert.
\end{equation}
On the other hand, $A^{\,l=3}_{\,m=3}$ is well modeled by
\begin{equation}\label{eq:42}
A^{l=3}_{m=3}(t,i)=b(\tau) + c(\tau)\vert \cos i \vert^{3/2},
\end{equation}
where  the  fitting  coefficients  $a,   b$  and  $c$  depend  on  the
interval of integration. 
Table~\ref{tab:4} shows the fitting  coefficients as a function of the
integration  time $\tau$.   From this  data it  is possible  to  fit a
function to establish the functional behavior of the coefficients with
respect   to  the   integration  time.    The  result   is   shown  in
Figure~\ref{fig:17}.   The  coefficients  $a,  b$  and  $c$  are  well
represented by
\begin{eqnarray}
a(\tau)&=&\alpha_1 \e^{- \tau^{\alpha_2}}, \label{eq:43}\\
b(\tau)&=&\beta_1 \e^{- \tau^{\beta_2}}, \label{eq:44}\\
c(\tau)&=&\gamma_1 \e^{- \tau^{\gamma_2}}, \label{eq:45}
\end{eqnarray}
where 
\begin{eqnarray}
\alpha_1&=&8.94  \pm 0.018, \label{eq:46}\\
\alpha_2&=&(8.352 \pm 0.0029)\times 10^{-2},\label{eq:47}\\
\beta_1&=&10.17 \pm 0.21, \label{eq:48}\\
\beta_2&=&(8.26 \pm 0.032)\times 10^{-2},\label{eq:49}\\ 
\gamma_1&=&5.90  \pm 0.033, \label{eq:50}\\
\gamma_2&=&(8.305  \pm 0.0084)\times 10^{-2}.\label{eq:51} 
\end{eqnarray}
The  asymptotic behavior of  the coefficients  suggests that  for long
integration times it is possible to consider them as constants.

\begin{table}[btp] 
  \begin{center}
    \caption{Fitting   coefficients   of   Eqns.~\eqref{eq:41}   and
      \eqref{eq:42}.   For  the  8  time intervals  we  compute  the
      fitting coefficients  $a, b$  and $c$. We  include the  error of
      each coefficient. }\label{tab:4}
    \begin{tabular*}{0.45\textwidth}{@{\extracolsep{\fill}}l|c|c|c}
      \hline \hline $\tau\;[\times 10^{7}]$ & $a(\tau)\;[\times 10^{-2}]$
      &      $b(\tau)\;[\times     10^{-2}]$     &
      $c(\tau)\;[\times 10^{-2}]$ \\ \hline
0.6047&	$22.45\pm0.066$&$26.99\pm0.116$&$15.21\pm0.171$\\
1.2093&	$18.04\pm0.035$&$21.57\pm0.062$&$12.28\pm0.092$\\
1.8140&	$15.76\pm0.025$&$18.90\pm0.046$&$10.75\pm0.067$\\
2.4186&	$14.28\pm0.019$&$17.19\pm0.036$&$ 9.76\pm0.054$\\
3.0233&	$13.21\pm0.016$&$15.99\pm0.031$&$ 9.03\pm0.046$\\
3.6279&	$12.38\pm0.014$&$15.07\pm0.027$&$ 8.47\pm0.041$\\
4.2326&	$11.72\pm0.012$&$14.33\pm0.025$&$ 8.00\pm0.037$\\
4.8372&	$11.18\pm0.011$&$13.73\pm0.024$&$ 7.62\pm0.035$\\
      \hline 
      \hline
    \end{tabular*}
  \end{center}
\end{table}

\begin{figure}[btp]
  \centering 
  \includegraphics[width=85mm]{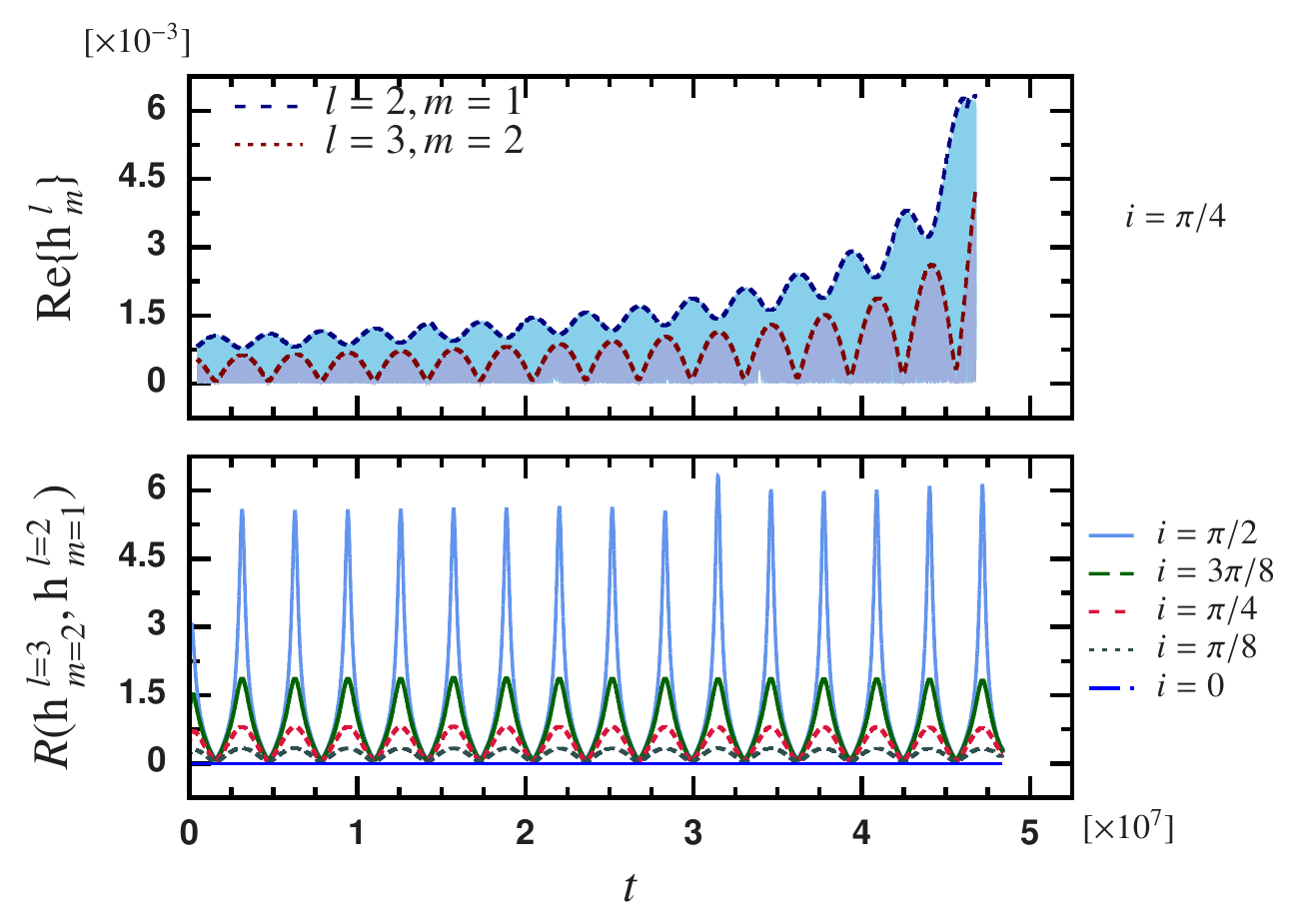}
  \caption{Variation of modes $l=3,m=2$ and $l=2,m=1$ as a function of
    the inclination angle $i$. The upper panel shows for $i=\pi/4$ the
    absolute  value of  the modes  and its  envelope. The  lower panel
    shows    the   quotient   of    the   envelopes    $l=3,m=2$   and
    $l=2,m=1$. Notice  that the resulting function  is almost periodic
    and does  not show the  characteristic growth close to  the merger
    phase.}
  \label{fig:18}
\end{figure}

Alternatively, it is possible to relate the inclination angle $i$ with
the   maximum  of   the  modes   $l=3,m=2$  and   $l=2,m=1$.    As  in
Sec.~\ref{sec:vari-eccentr-extern},  we compute  the  envelope of  the
modes using  a low-pass filter.  The upper  panel in Figure~\ref{fig:18}
shows  the  result for  the  angle  $i=\pi/4$.   The quotient  of  the
envelope  of  the  modes  $l=3,m=2$  and $l=2,m=1$  gives  a  periodic
function which  removes the  growth of the  modes close to  the merger
time.  We  define  the   function  $R$  which  \textit{rectifies}  the
envelopes as
\begin{equation}\label{eq:52}
R(\mathrm{h}^{l=3}_{m=2},\mathrm{h}^{l=2}_{m=1}):= \frac{\mathrm{Env}[\mathrm{Re}\{\mathrm{h}^{l=3}_{m=2}\}]}{\mathrm{Env}[\mathrm{Re}\{\mathrm{h}^{l=2}_{m=1}\}]}.
\end{equation}
The  lower panel  of Figure~\ref{fig:18}  shows the  result  of applying
\eqref{eq:52} to  our data. Notice that  in the case  of $i=\pi/2$ the
values after $t=3\times  10^7$ are a little erratic.  For our analysis
we consider for $i=\pi/2$ only the points before $t=3\times 10^7$.

\begin{figure}[btp]
  \centering 
  \includegraphics[width=85mm]{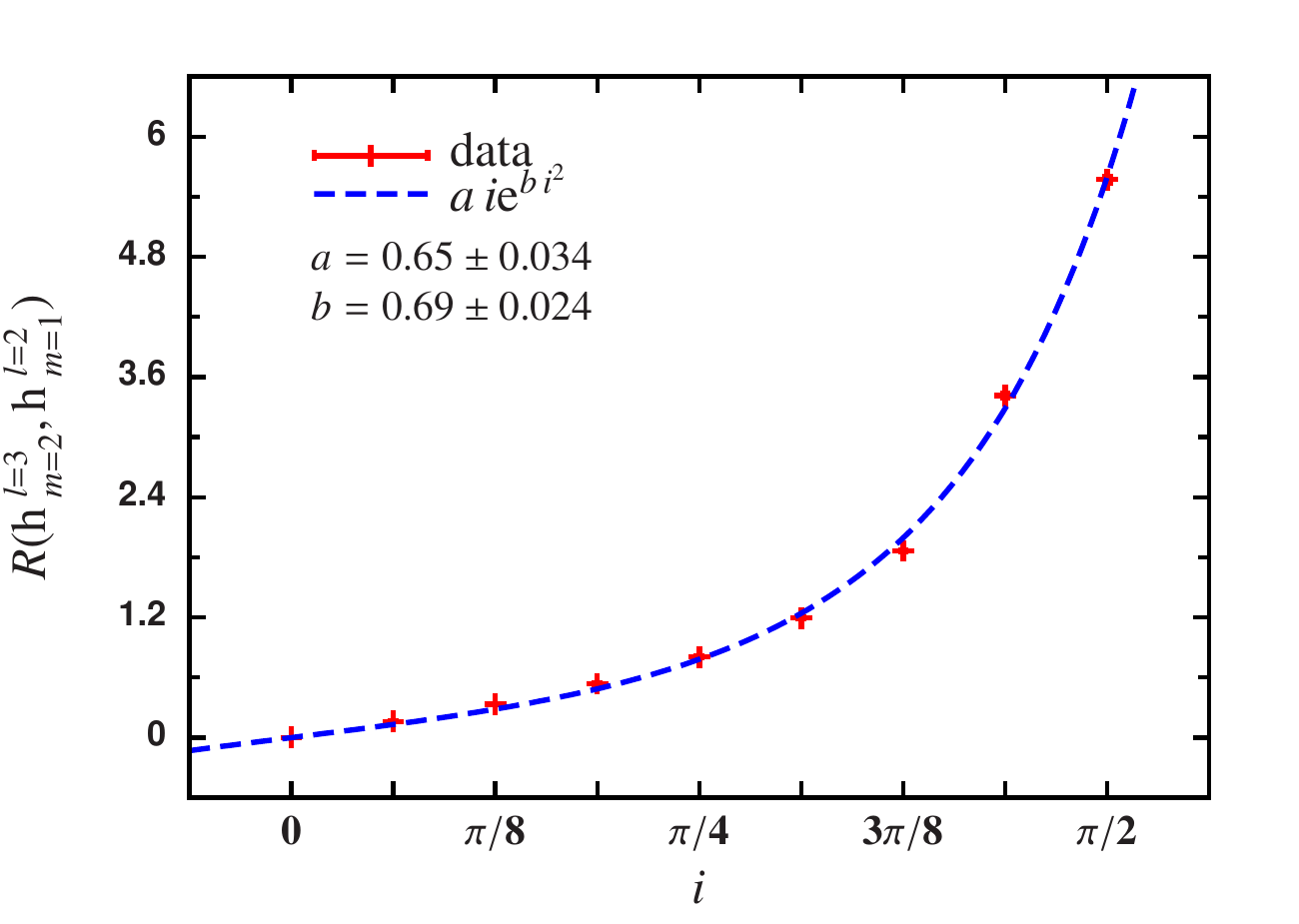}
  \caption{The maximum of \eqref{eq:52} as function of the inclination
    angle  $i$.  The functional  behavior is  well represented  by the
    function $a\,i \e^{b\,i^2}$. }
  \label{fig:19}
\end{figure}

From the resulting function we compute numerically the local maxima of
\eqref{eq:52}. Table~\ref{tab:5} shows the result. For this purpose we
perform  additional   simulations  for  angles   $\pi/16$,  $3\pi/16$,
$5\pi/16$ and $7\pi/16$.
We  fit  to  the  data  the function  $f(i)=a\,i  \e^{b\,i^2}$,  where
$a=0.65\pm0.034$ and  $b=0.69\pm0.024$.  Figure~\ref{fig:19} shows the
result, notice that the functional behavior is well represented by the
fitted function.
\begin{table}[btp] 
  \begin{center}
    \caption{The  maximum  of  \eqref{eq:52}  as  a  function  of  the
      inclination  angle  $i$. Listed  is  the  average  value of  the
      maxima, while  the error is  given by the standard  deviation of
      the data.}\label{tab:5}
    \begin{tabular*}{0.35\textwidth}{@{\extracolsep{\fill}}r|c|c}
      \hline \hline $i$ & 
      $\mathrm{Max}[R(\mathrm{h}^{\;l=3}_{\;m=2},\mathrm{h}^{\;l=2}_{\;m=1})]$&
      Variation (\%)  \\ \hline
      $0$      &$0$&	0\\
      $\pi/16$ &$0.1608\pm0.00077$&	0.48\\
      $\pi/8$  &$0.335\pm0.0012$  &	0.36\\
      $3\pi/16$&$0.538\pm0.0029$  &	0.54\\
      $\pi/4$  &$0.806\pm0.0049$  &	0.61\\
      $5\pi/16$&$1.193\pm0.0056$  &	0.47\\
      $3\pi/8$ &$1.864\pm0.0093$  &	0.50\\
      $7\pi/16$&$3.41\pm0.026$    &	0.75\\
      $\pi/2$  &$5.57\pm0.033$    &	0.60\\
      \hline 
      \hline
    \end{tabular*}
  \end{center}
\end{table}

In both cases, using the relative ``area'' of the modes or the maximum
of the  ``rectified'' modes,  we obtain quite  a simple  behavior. The
advantage  of the  second method  is that  it does  not depend  on the
integration time $\tau$.

\subsubsection{Initial separation of the external binary}
\label{sec:init-separ-extern}
The last  numerical experiment examines the dependence  on the initial
separation of the external binary $r_3$.  We set the value of $r_3$ to
312.5, 625, 1250, 2500, 5000,  and 10000. For $r_3=312.5$ the external
body  is  ejected  from the  binary  after  a  few orbits,  the  other
configurations are stable.

\begin{figure}[btp]
  \centering 
  \includegraphics[width=85mm]{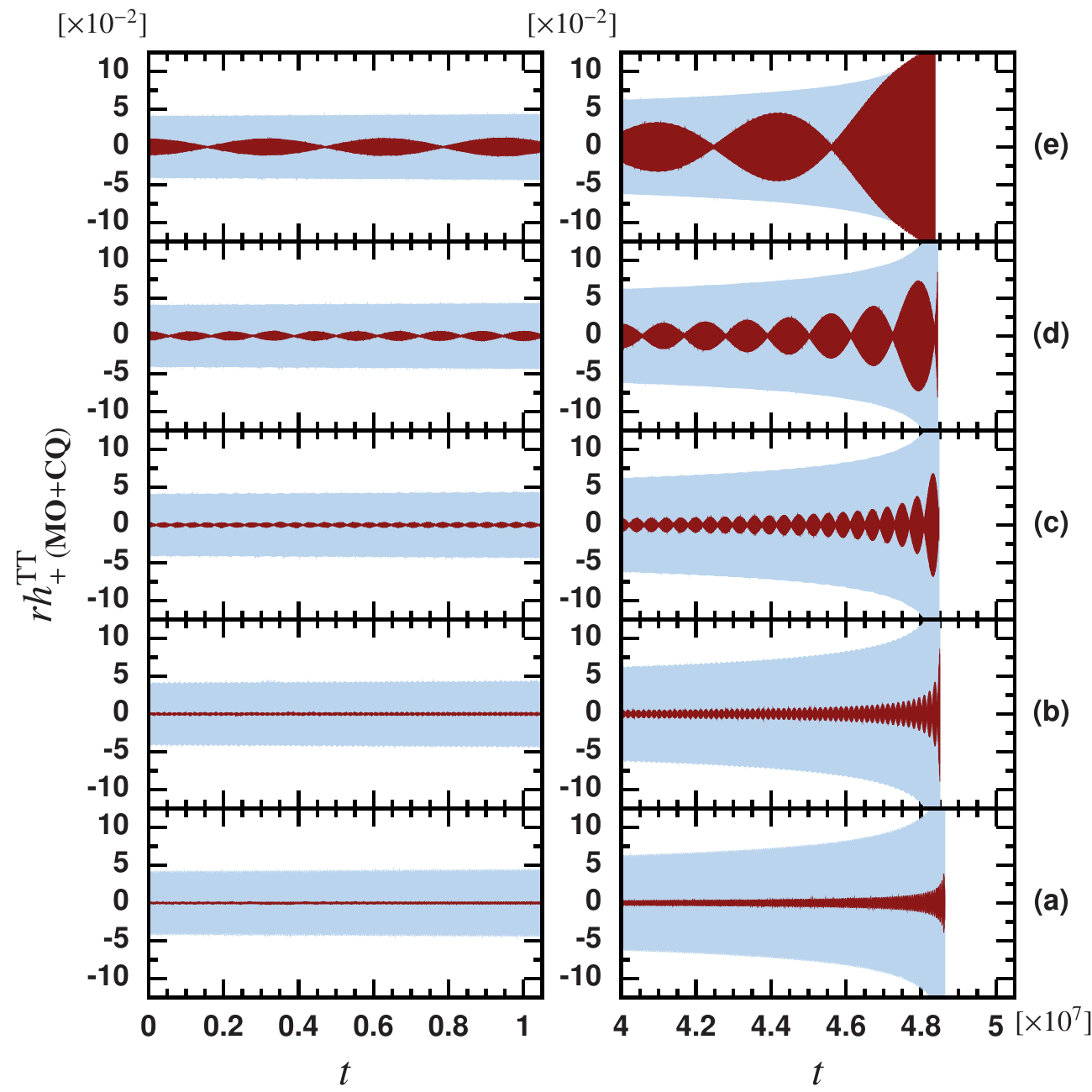}
  \caption{Planar  hierarchical   system.   Modulation  of   the  mass
    octupole plus  the current quadrupole  as function of  the initial
    separation  of the  external body.   The initial  separation $r_3$
    takes  the  values   625  \textbf{(a)},  1250  \textbf{(b)},  2500
    \textbf{(c)}, 5000  \textbf{(d)} and 10000  \textbf{(e)}. Shown on
    the left is the evolution  for $t\in[0,10^7]$ and on the right for
    $t\in[4\times 10^7,5\times 10^7]$.}
  \label{fig:20}
\end{figure}

Figure~\ref{fig:20}  shows the  sum  of the  mass  octupole and  current
quadrupole  contributions  to  the  waveform.  The  frequency  of  the
modulation of the waveform increases when the separation and hence the
orbital period of the external  binary is decreased.  One orbit of the
external binary  corresponds to the time  between two of  the nodes of
the  mass  octupole  plus  current quadrupole  contribution  shown  in
Figure~\ref{fig:20}.   The influence  of  a third  body  is not  clearly
defined when the period of the external binary is similar to the inner
binary.   For  small separations,  on  the  scale  shown there  is  no
modulation  of the waves  visible (see  Figure~\ref{fig:20} \textbf{(a)}
and \textbf{(b)}).  When the initial separation of the external binary
is increased, at some distance most  of the inspiral and merger of the
inner binary happens before the external binary completes one orbit.

\section{DISCUSSION}
\label{sec:discussion}

We   performed   post-Newtonian  simulations   for   a  selection   of
hierarchical configurations as an example for a three-body system, and
we analyzed  the waveforms. Based  on these simulations we  examined a
number of different physical aspects of the system.

First of all, looking at the mass octupole and current quadrupole part
of  the  waveform,  it  is  possible to  distinguish  between  such  a
hierarchical (also called Jacobian) triple system and a binary system,
an issue that has been discussed in \cite{TorHatAsa09,Asa09}.

In terms of  the merger time, the perturbed  binary merges later.  For
mass ratio 10:20:1, the delay of  the merger is $27\%$ compared to the
binary with 10:20, which is perhaps surprisingly large.
However, let  us note that  even a small  perturbation due to  a third
object can have a large  effect when integrated over about 4000 orbits
of the  inner binary (i.e.\  there is less  than a $0.01\%$  delay per
orbit).
As  we  have  shown,  the  delay  depends  only  very  weakly  on  the
inclination   angle  or   the  distance   to  the   third   body,  see
Figs.~\ref{fig:15} and  \ref{fig:20}.  This may be  expected since the
force due to the third  body periodically increases but also decreases
the force  between the objects of  the inner binary  (depending on the
orientation  of the  binary with  respect  to the  third body),  which
apparently averages out over several orbits of the inner binary.
As a  cross check we also  performed simulations where  the third mass
approaches zero, and  in this case the merger  time does approach that
of the binary.

As far as the approximation method is concerned, we find that there is
a  significant  difference in  the  merger  time  for a  system  which
includes  Newtonian  dynamics and  2.5~PN  radiation  compared to  the
inclusion of 1~PN or 2~PN corrections to the dynamics.
The  inclusion of  1~PN corrections  to the  conservative part  of the
Hamiltonian produces  a change of  16\% in the merger  time.  However,
the  inclusion  of  2~PN  corrections  does  not  make  a  significant
difference  to either  the waveform  or the  merger time  (only around
0.5\%).

The variation  of the eccentricity  of the external binary  shows that
the  period of  the  third body  is  well described  by the  Newtonian
dynamics. From the modulation of the waveform modes (particularly from
the  sum  of the  $l=2,m=1$  and $l=m=3$  modes),  it  is possible  to
distinguish two frequencies which  are related to the eccentricity via
a Newtonian expression.

We  established a  link between  the  amplitude of  the $l=3,m=2$  and
$l=m=3$ modes and the angle of the osculating orbital planes. In order
to  extract   the  information  given   by  the  waves  we   used  two
methods. First, we used the relative area covered by the $l=3,m=2$ and
$l=m=3$ modes with respect to the area covered by the mode $l=m=2$. In
this  case the  contribution  of the  $l=3,m=2$  mode is  particularly
simple. It is zero for planar  motion and increases as a sine function
of the inclination  angle. The second method is  based on the quotient
of  the  envelope  of the  $l=3,m=2$  mode  and  the envelope  of  the
$l=2,m=1$ mode.   The resulting function  is almost periodic  and does
not contain  the characteristic growth  of the waveforms close  to the
merger phase. In  this case, it is possible  to relate the inclination
angle to the amplitude of the resulting function.
The  modulation  produced  by  the  third  body  on  the  $l=3$  modes
characterizes  the  period  of  the external  binary.  Decreasing  the
initial separation  of the external  body produces a  higher frequency
modulation, until it  is no longer possible to  discern a well defined
modulation of  the waveform.   In our simulations,  when there  are no
well defined internal and external binaries the system is not stable.

Our results  provide additional evidence to a  conjecture first stated
in \cite{TorHatAsa09}, that  in order to characterize a  system of $n$
compact  objects,  it is  necessary  to  perform  an analysis  of  the
waveform which includes at least the $l\leq n$ modes.  As we showed in
the last  numerical experiment,  when the third  body is close  to the
binary it  is not  evident how to  extract information related  to the
dynamics of a  particular body. It is necessary  to perform a detailed
study  of  nonhierarchical  triple  systems  to  determine  how  much
information we can extract from more general cases.
More detailed statements based on the higher modes of the waveform are
possible   but   require   an   extensive  parameter   study.    Other
configurations include for example a massive compact object perturbing
a binary, or the scattering and  capture of a third body.  The present
examples showed  the type of  characterization that are  possible with
the techniques developed above.

As a final  comment, let us point out that  chaotic behavior of triple
systems is well known in the Newtonian case (see e.g.\ \cite{ValKar06}
and references therein). For binaries,  it is known that chaos appears
when using  certain post-Newtonian approximations for  systems of {\em
  spinning}              binaries              (see              e.g.\
\cite{Lev00,CorLev03,BarLev03,Lev06,GopKni05,Han08,SchRas01}).   As  a
natural  generalization  of the  Newtonian  case  we  expect that  the
three-body problem  exhibits chaotic  behavior as well.   An important
question is,  how does the emission of  gravitational radiation change
the chaotic  properties of the system?   We consider this  a topic for
future study.

\acknowledgments

It  is a  pleasure to  thank Gerhard  Sch\"afer,  Sebastiano Bernuzzi,
David Hilditch, and Milton  Ruiz for valuable discussions and comments
on  the manuscript.  This  work was  supported  in part  by DFG  grant
SFB/Transregio 7 and by DLR grant LISA Germany.

\appendix

\bigskip

\section{First and second post-Newtonian Hamiltonian}
\label{sec:first-second-post}

Here  we   reproduce  in  our   notation  the  Hamiltonian   given  in
\cite{Sch87}, with some factorizations and changes in the summation of
the terms  T1 and T2, which  are marked by braces  below.  Our version
(worked   out  with   G.   Sch\"afer)   fixes  the   typos   noted  in
\cite{LouNak08},  giving a formula  equivalent to  \cite{LouNak08} but
written in a different way.  The issue is how the four-point functions
of  \cite{OhtKimHii75}  are reduced  to  explicit  triple  sums for  a
three-body problem.  The  first and second post-Newtonian Hamiltonians
are

\begin{equation}\label{eq:53}
\begin{split}
H_{1} =& -\frac{1}{8}\sum_a m_a \left(\frac{\vec{p}_a^{\;2}}{m_a^2}
\right )^2 - \frac{1}{4} \sum_a \sum_{b\neq a} \frac{1}{r_{ab}} \Big (
6 \frac{m_b}{m_a} \vec{p}_a^{\;2}  \\ &  - 7\vec{p}_a \cdot \vec{p}_b -
(\hat{n}_{ab} \cdot \vec{p}_a)(\hat{n}_{ab} \cdot \vec{p}_b)   \Big ) \\
&+\frac{1}{2}\sum_{a}\sum_{b\neq a}\sum_{c\neq a} \frac{m_a m_b m_c}{r_{ab}r_{ac}},
\end{split} 
\end{equation}

\begin{widetext}
  \begin{eqnarray}
    H_{2}
    &=&
    \frac{1}{16} \sum_{a} m_a \left(\frac{\vec{p}_a^{\;2}}{m_a^2}\right)^3 
    +
    \frac{1}{16} \sum_{a}\sum_{b \neq a} 
    \frac{m_a^{-1} m_b^{-1}}{r_{ab}}
    \Big[
    10 \left(\frac{m_b}{m_a} \vec{p}_a^{\;2} \right)^2
    - 11 \vec{p}_a^{\;2} \vec{p}_b^{\;2}
    - 2 \left( \vec{p}_a \cdot \vec{p}_b \right)^2
    \nonumber \\
    &&
    + 10 \vec{p}_a^{\;2} \left( \hat{n}_{ab} \cdot \vec{p}_b \right)^2
    - 12 \left( \vec{p}_a \cdot \vec{p}_b \right)
    \left( \hat{n}_{ab} \cdot \vec{p}_a \right)
    \left( \hat{n}_{ab} \cdot \vec{p}_b \right)
    - 3 \left( \hat{n}_{ab} \cdot \vec{p}_a \right)^2
    \left( \hat{n}_{ab} \cdot \vec{p}_b \right)^2
    \Big]
    \nonumber \\ && 
    +
    \frac{1}{8} \sum_{a}\sum_{b\ne a}\sum_{c\ne a}  
    \frac{1}{r_{ab}\,r_{ac}}
    \Big[
    18 \frac{m_b m_c}{m_a}\vec{p}_a^{\;2} 
    +14 \frac{m_a m_c}{m_b} \vec{p}_b^{\;2} 
    -2 \frac{m_a m_c}{m_b}\left( \hat{n}_{ab}\cdot \vec{p}_{b} \right)^2
    \nonumber \\
    && 
    -50 m_c (\vec{p}_{a}\cdot \vec{p}_{b})
    +17 m_a (\vec{p}_{b}\cdot \vec{p}_{c})
    - 14 m_c \left( \hat{n}_{ab}\cdot \vec{p}_{a} \right)
    \left( \hat{n}_{ab}\cdot \vec{p}_{b} \right)
    \nonumber \\
    && 
    + 14 m_a \left( \hat{n}_{ab}\cdot \vec{p}_{b} \right)
    \left( \hat{n}_{ab}\cdot \vec{p}_{c} \right)
    + m_a (\hat{n}_{ab}\cdot \hat{n}_{ac}) 
    \left( \hat{n}_{ab}\cdot \vec{p}_{b} \right)
    \left( \hat{n}_{ac}\cdot \vec{p}_{c} \right)
    \Big]
    \nonumber \\
    && 
    +
    \frac{1}{8} \sum_{a}\sum_{b\ne a}\sum_{c\ne a}  
    \frac{1}{r_{ab}^2}
    \Big[
    2 m_b \left( \hat{n}_{ab}\cdot \vec{p}_{a} \right)
    \left( \hat{n}_{ac}\cdot \vec{p}_{c} \right)
    + 2 m_b \left( \hat{n}_{ab}\cdot \vec{p}_{b} \right)
    \left( \hat{n}_{ac}\cdot \vec{p}_{c} \right)
    \nonumber \\
    && 
    +\frac{m_a m_b}{m_c} \left( 5 (\hat{n}_{ab}\cdot \hat{n}_{ac})\vec{p}_c^{\;2}
      - (\hat{n}_{ab}\cdot \hat{n}_{ac}) 
      \left( \hat{n}_{ac}\cdot \vec{p}_{c} \right)^2 
      - 14 \left( \hat{n}_{ab}\cdot \vec{p}_{c} \right)
      \left( \hat{n}_{ac}\cdot \vec{p}_{c} \right) \right)
    \Big]
    \nonumber \\
    && 
    + 
    \frac{1}{4} \sum_{a}\sum_{ b\ne a}
    \frac{m_a}{r_{ab}^2}
    \Big[
    \frac{m_b}{m_a}\vec{p}_a^{\;2} + \frac{m_a}{m_b}\vec{p}_b^{\;2} 
    - 2 (\vec{p}_a \cdot \vec{p}_b )
    \Big]
    \nonumber \\
    && 
    +
    \frac{1}{2} \sum_{a}\sum_{ b\ne a} \sum_{c\ne a,b}
    \frac{( n^i_{ab} + n^i_{ac} ) ( n^j_{ab} + n^j_{cb} )}{\left(r_{ab} + r_{bc} + r_{ca}\right)^2}   
    \Big[  
    8 m_b (p_{ai}p_{cj})
    -16 m_b (p_{aj}p_{ci})
    \nonumber \\
    && 
    +3 m_c (p_{ai}p_{bj})
    +4 \frac{m_a m_b}{m_c} (p_{ci}p_{cj})
    + \frac{m_b m_c}{m_a}( p_{ai}p_{aj})
    \Big]
    \nonumber \\
    && 
    +
    \frac{1}{2} \sum_{a}\sum_{ b\ne a}\sum_{ c\ne a,b}
    \frac{m_a m_b m_c}{\left(r_{ab} + r_{bc} + r_{ca}\right) r_{ab}}
    \Big[
    8 \frac{\vec{p}_{a} \cdot \vec{p}_{c} 
      -\left( \hat{n}_{ab} \cdot \vec{p}_a \right)
      \left( \hat{n}_{ab} \cdot \vec{p}_c \right)}{m_a m_c}
    \nonumber \\
    && 
    -3 \frac{\vec{p}_{a} \cdot \vec{p}_{b}
      -\left( \hat{n}_{ab} \cdot \vec{p}_a \right)
      \left( \hat{n}_{ab} \cdot \vec{p}_b \right)}{m_a m_b}
    -4 \frac{\vec{p}_c^{\;2} - \left( \hat{n}_{ab} \cdot \vec{p}_c \right)^2}{m_c^2}
    -\frac{\vec{p}_a^{\;2} - \left( \hat{n}_{ab} \cdot \vec{p}_a \right)^2}{m_a^2}
    \Big]
    \nonumber \\
    && 
    -\overbrace{  \frac{1}{2} \sum_{a}\sum_{b\ne a} \left ( \sum_{ c \ne a,b}
        \frac{m_a^2 m_b m_c }{r_{ab}^2\, r_{bc}} + \frac{1}{2}\sum_{ c \ne b}
        \frac{m_a^2 m_b m_c }{r_{ab}^2\, r_{bc}} \right )}^{\mathrm{T1}}
    \nonumber \\ 
    && 
    -\overbrace{\frac{3}{8} \sum_{a} \sum_{b\ne a} \left ( \sum_{ c\ne a}
      \frac{m_a^2 m_b m_c}{r_{ab}^2\, r_{ac}}+\sum_{ c\ne a,b}
      \frac{m_a^2 m_b m_c}{r_{ab}^2\, r_{ac}} \right)}^{\mathrm{T2}}
    \nonumber \\ 
    && 
    -\frac{3}{8} \sum_{a} \sum_{b\ne a} \sum_{ c\ne a,b}
    \frac{m_a^2 m_b m_c}{r_{ab}^2\, r_{ac}\,r_{bc}} 
    \nonumber \\ 
    && 
    -\frac{1}{64}\sum_{a} \sum_{ b\ne a} \sum_{ c\ne a,b}
    \frac{m_a^2 m_b m_c}{r_{ab}\, r_{ac}^3\,r_{bc}}
    \big\{
    18r_{ac}^2-60 r_{bc}^2-24 r_{ac}(r_{ab}+r_{bc})
    \nonumber \\ &&
    +60 \frac{r_{ac}r_{bc}^2}{r_{ab}}+56 r_{ab} r_{bc}
    -72 \frac{r_{bc}^3}{r_{ab}}+35 \frac{r_{bc}^4}{r_{ab}^2}+6r_{ab}^2
    \big\}
    - \frac{1}{4} \sum_{a} \sum_{ b\ne a}  
    \frac{m_a^2 m_b^2}{r_{ab}^3} \,. \label{eq:54}
\end{eqnarray}
\end{widetext}

\section{Lagrange triangle solution waveform}
\label{sec:lagr-triangle-solut}

Here  we  summarize the  expressions  for  the  mass quadrupole,  mass
octupole, and  current quadrupole  waveforms for each  polarization of
the Lagrange  triangle solution.  See \cite{Asa09} for  details on the
calculation     of     this      expression.      We     denote     by
$a:=r_{12}=r_{13}=r_{23}$ the separation  between each pair of bodies.
$m_1$,  $m_2$  and  $m_3$   are  the  dimensionless  mass  parameters,
$\omega=a^{-3/2}$ is  the orbital frequency, $r$ is  the distance from
the  observer  to  the   source  and  $\theta$  is  the  observational
direction.  We define the following auxiliary quantities:
\begin{eqnarray}
\mu_i&:=& \sqrt{m_j^2+m_j m_k +m_k^2} , \label{eq:55}\\ 
\phi_1&:=& 0, \label{eq:56} 
\end{eqnarray}

\begin{eqnarray}
\phi_2&:=& \arccos \left( \frac{\mu_1^2+\mu_3^2-1}{2 \mu_1 \mu_3} \right), \label{eq:57}\\
\phi_3&:=& -\arccos \left( \frac{\mu_1^2+\mu_2^2-1}{2 \mu_1 \mu_2} \right), \label{eq:58} 
\end{eqnarray}
where $j \neq  i, \; k \neq i,j$. The plus  and cross polarizations of
the mass quadrupole waveform are
\begin{equation}
r h_{+}^{MQ} = -(3+\cos 2 \theta) a^2 \omega^2 \sum_{i=1}^3 m_i \mu_i^2 \cos (2 (\omega t + \phi_i ) ), \label{eq:59}
\end{equation}

\begin{equation}
r h_{\times}^{MQ} = -4 \cos \theta a^2 \omega^2 \sum_{i=1}^3 m_i \mu_i^2 \sin (2 (\omega t + \phi_i ) ), \label{eq:60}
\end{equation}
the expressions for the current quadrupole are 
\begin{equation}
r h_{+}^{CQ}=\frac{4 a^3 \omega^3}{3} \sin \theta \sum_{i=1}^3  m_i \mu_i^3 \cos (\omega t+\phi_i),\label{eq:61} 
\end{equation}
\begin{equation}
r h_{\times}^{CQ}=\frac{2 a^3 \omega^3}{3} \sin (2\theta) \sum_{i=1}^3  m_i \mu_i^3 \sin (\omega t+\phi_i),\label{eq:62}
\end{equation}
and the waveforms for the mass octupole are given by 
\begin{equation}
\begin{split}
r h_{+}^{MO}=&\frac{a^3 \omega^3}{12} \sin \theta  \sum_{i=1}^3  m_i \mu_i^3 \left[(3 \cos^2 \theta - 1)  \cos ( \omega t + \phi_i ) \right.\\
&\left. -(27 (1+\cos^2 \theta)) \cos (3 (\omega t  + \phi_i)) \right] \label{eq:63} 
\end{split}
\end{equation}
\begin{equation}
\begin{split}
r h_{\times}^{MO}=&\frac{a^3 \omega^3}{12} \sin (2\theta) \sum_{i=1}^3  m_i \mu_i^3 \left[ \sin (\omega t+\phi_i) \right.\\
&\left.- 27 \sin(3 (\omega t+\phi_i))  \right]\label{eq:64}
\end{split}
\end{equation}


\bibliography{../Tex/refs_extra}

\begin{thebibliography}{10}%
\makeatletter
\providecommand \@ifxundefined [1]{%
 \ifx #1\undefined \expandafter \@firstoftwo
 \else \expandafter \@secondoftwo
\fi
}%
\providecommand \@ifnum [1]{%
 \ifnum #1\expandafter \@firstoftwo
 \else \expandafter \@secondoftwo
\fi
}%
\providecommand \enquote [1]{``#1''}%
\providecommand \bibnamefont  [1]{#1}%
\providecommand \bibfnamefont [1]{#1}%
\providecommand \citenamefont [1]{#1}%
\providecommand\href[0]{\@sanitize\@href}%
\providecommand\@href[1]{\endgroup\@@startlink{#1}\endgroup\@@href}%
\providecommand\@@href[1]{#1\@@endlink}%
\providecommand \@sanitize [0]{\begingroup\catcode`\&12\catcode`\#12\relax}%
\@ifxundefined \pdfoutput {\@firstoftwo}{%
 \@ifnum{\z@=\pdfoutput}{\@firstoftwo}{\@secondoftwo}%
}{%
 \providecommand\@@startlink[1]{\leavevmode\special{html:<a href="#1">}}%
 \providecommand\@@endlink[0]{\special{html:</a>}}%
}{%
 \providecommand\@@startlink[1]{%
  \leavevmode
  \pdfstartlink
   attr{/Border[0 0 1 ]/H/I/C[0 1 1]}%
   user{/Subtype/Link/A<</Type/Action/S/URI/URI(#1)>>}%
  \relax
 }%
 \providecommand\@@endlink[0]{\pdfendlink}%
}%
\providecommand \url  [0]{\begingroup\@sanitize \@url }%
\providecommand \@url [1]{\endgroup\@href {#1}{\urlprefix}}%
\providecommand \urlprefix [0]{URL }%
\providecommand \Eprint[0]{\href }%
\@ifxundefined \urlstyle {%
  \providecommand \doi [1]{doi:\discretionary{}{}{}#1}%
}{%
  \providecommand \doi [0]{doi:\discretionary{}{}{}\begingroup
  \urlstyle{rm}\Url }%
}%
\providecommand \doibase [0]{http://dx.doi.org/}%
\providecommand \Doi[1]{\href{\doibase#1}}%
\providecommand \bibAnnote [3]{%
  \BibitemShut{#1}%
  \begin{quotation}\noindent
    \textsc{Key:}\ #2\\\textsc{Annotation:}\ #3%
  \end{quotation}%
}%
\providecommand \bibAnnoteFile [2]{%
  \IfFileExists{#2}{\bibAnnote {#1} {#2} {\input{#2}}}{}%
}%
\providecommand \typeout [0]{\immediate \write \m@ne }%
\providecommand \selectlanguage [0]{\@gobble}%
\providecommand \bibinfo [0]{\@secondoftwo}%
\providecommand \bibfield [0]{\@secondoftwo}%
\providecommand \translation [1]{[#1]}%
\providecommand \BibitemOpen[0]{}%
\providecommand \bibitemStop [0]{}%
\providecommand \bibitemNoStop [0]{.\EOS\space}%
\providecommand \EOS [0]{\spacefactor3000\relax}%
\providecommand \BibitemShut [1]{\csname bibitem#1\endcsname}%
\bibitem{ChiImaAsa07}%
  \BibitemOpen
  \bibfield{author}{%
  \bibinfo {author} {\bibfnamefont{T.}~\bibnamefont{Chiba}}, \bibinfo {author}
  {\bibfnamefont{T.}~\bibnamefont{Imai}},\ and\ \bibinfo {author}
  {\bibfnamefont{H.}~\bibnamefont{Asada}},\ }%
  \bibfield{journal}{%
  \Doi{10.1111/j.1365-2966.2007.11619.x}{\bibinfo {journal} {Mon. Not. R.
  Astron. Soc.}}\ }%
  \textbf{\bibinfo {volume} {377}},\ \bibinfo {pages} {269} (\bibinfo {year}
  {2007}),\
  \Eprint{http://arxiv.org/abs/astro-ph/0609773}{arXiv:astro-ph/0609773}%
  \bibAnnoteFile{NoStop}{ChiImaAsa07}%
\bibitem{TorHatAsa09}%
  \BibitemOpen
  \bibfield{author}{%
  \bibinfo {author} {\bibfnamefont{Y.}~\bibnamefont{Torigoe}}, \bibinfo
  {author} {\bibfnamefont{K.}~\bibnamefont{Hattori}},\ and\ \bibinfo {author}
  {\bibfnamefont{H.}~\bibnamefont{Asada}},\ }%
  \bibfield{journal}{%
  \Doi{10.1103/PhysRevLett.102.251101}{\bibinfo {journal} {Phys. Rev. Lett.}}\
  }%
  \textbf{\bibinfo {volume} {102}},\ \bibinfo {pages} {251101} (\bibinfo
  {month} {Jun}\ \bibinfo {year} {2009}),\
  \Eprint{http://arxiv.org/abs/0906.1448}{arXiv:0906.1448 [gr-qc]}%
  \bibAnnoteFile{NoStop}{TorHatAsa09}%
\bibitem{Asa09}%
  \BibitemOpen
  \bibfield{author}{%
  \bibinfo {author} {\bibfnamefont{H.}~\bibnamefont{Asada}},\ }%
  \bibfield{journal}{%
  \Doi{10.1103/PhysRevD.80.064021}{\bibinfo {journal} {Phys. Rev. D}}\ }%
  \textbf{\bibinfo {volume} {80}},\ \bibinfo {pages} {064021} (\bibinfo {month}
  {Sep}\ \bibinfo {year} {2009}),\
  \Eprint{http://arxiv.org/abs/0907.5091}{arXiv:0907.5091 [gr-qc]}%
  \bibAnnoteFile{NoStop}{Asa09}%
\bibitem{ValKar06}%
  \BibitemOpen
  \bibfield{author}{%
  \bibinfo {author} {\bibfnamefont{M.~J.}\ \bibnamefont{Valtonen}}\ and\
  \bibinfo {author} {\bibfnamefont{H.}~\bibnamefont{Karttunen}},\ }%
  \emph{\bibinfo {title} {The three-body problem}}\ (\bibinfo {publisher}
  {Cambridge University Press},\ \bibinfo {address} {New York},\ \bibinfo
  {year} {2006})\ ISBN \bibinfo {isbn} {0-521-85224-2 (hardcover)}%
  \bibAnnoteFile{NoStop}{ValKar06}%
\bibitem{GulMilHam03}%
  \BibitemOpen
  \bibfield{author}{%
  \bibinfo {author} {\bibfnamefont{K.}~\bibnamefont{Gultekin}}, \bibinfo
  {author} {\bibfnamefont{M.~C.}\ \bibnamefont{Miller}},\ and\ \bibinfo
  {author} {\bibfnamefont{D.~P.}\ \bibnamefont{Hamilton}},\ }%
  \bibfield{journal}{%
  \Doi{10.1063/1.1629425}{\bibinfo {journal} {AIP Conf. Proc.}}\ }%
  \textbf{\bibinfo {volume} {686}},\ \bibinfo {pages} {135} (\bibinfo {year}
  {2003}),\
  \Eprint{http://arxiv.org/abs/astro-ph/0306204}{arXiv:astro-ph/0306204}%
  \bibAnnoteFile{NoStop}{GulMilHam03}%
\bibitem{GulMilHam04}%
  \BibitemOpen
  \bibfield{author}{%
  \bibinfo {author} {\bibfnamefont{K.}~\bibnamefont{Gultekin}}, \bibinfo
  {author} {\bibfnamefont{M.~C.}\ \bibnamefont{Miller}},\ and\ \bibinfo
  {author} {\bibfnamefont{D.~P.}\ \bibnamefont{Hamilton}},\ }%
  \bibfield{journal}{%
  \Doi{10.1086/424809}{\bibinfo {journal} {The Astrophysical Journal}}\ }%
  \textbf{\bibinfo {volume} {616}},\ \bibinfo {pages} {221} (\bibinfo {year}
  {2004}),\
  \Eprint{http://arxiv.org/abs/astro-ph/0402532}{arXiv:astro-ph/0402532}%
  \bibAnnoteFile{NoStop}{GulMilHam04}%
\bibitem{GulMilHam05a}%
  \BibitemOpen
  \bibfield{author}{%
  \bibinfo {author} {\bibfnamefont{K.}~\bibnamefont{Gultekin}}, \bibinfo
  {author} {\bibfnamefont{M.~C.}\ \bibnamefont{Miller}},\ and\ \bibinfo
  {author} {\bibfnamefont{D.~P.}\ \bibnamefont{Hamilton}},\ }%
  \bibfield{journal}{%
  \Doi{10.1086/499917}{\bibinfo {journal} {The Astrophysical Journal}}\ }%
  \textbf{\bibinfo {volume} {640}},\ \bibinfo {pages} {156} (\bibinfo {year}
  {2006}),\
  \Eprint{http://arxiv.org/abs/astro-ph/0509885}{arXiv:astro-ph/0509885}%
  \bibAnnoteFile{NoStop}{GulMilHam05a}%
\bibitem{IwaFunMak06}%
  \BibitemOpen
  \bibfield{author}{%
  \bibinfo {author} {\bibfnamefont{M.}~\bibnamefont{Iwasawa}}, \bibinfo
  {author} {\bibfnamefont{Y.}~\bibnamefont{Funato}},\ and\ \bibinfo {author}
  {\bibfnamefont{J.}~\bibnamefont{Makino}},\ }%
  \bibfield{journal}{%
  \Doi{10.1086/507473}{\bibinfo {journal} {The Astrophysical Journal}}\ }%
  \textbf{\bibinfo {volume} {651}},\ \bibinfo {pages} {1059} (\bibinfo {year}
  {2006}),\
  \Eprint{http://arxiv.org/abs/astro-ph/0511391}{arXiv:astro-ph/0511391}%
  \bibAnnoteFile{NoStop}{IwaFunMak06}%
\bibitem{HofLoe07}%
  \BibitemOpen
  \bibfield{author}{%
  \bibinfo {author} {\bibfnamefont{L.}~\bibnamefont{Hoffman}}\ and\ \bibinfo
  {author} {\bibfnamefont{A.}~\bibnamefont{Loeb}},\ }%
  \bibfield{journal}{%
  \Doi{10.1111/j.1365-2966.2007.11694.x}{\bibinfo {journal} {Mon. Not. R.
  Astron. Soc.}}\ }%
  \textbf{\bibinfo {volume} {377}},\ \bibinfo {pages} {957} (\bibinfo {year}
  {2007}),\
  \Eprint{http://arxiv.org/abs/astro-ph/0612517}{arXiv:astro-ph/0612517}%
  \bibAnnoteFile{NoStop}{HofLoe07}%
\bibitem{Hofloe06}%
  \BibitemOpen
  \bibfield{author}{%
  \bibinfo {author} {\bibfnamefont{L.}~\bibnamefont{Hoffman}}\ and\ \bibinfo
  {author} {\bibfnamefont{A.}~\bibnamefont{Loeb}},\ }%
  \bibfield{journal}{%
  \Doi{10.1086/501230}{\bibinfo {journal} {The Astrophysical Journal}}\ }%
  \textbf{\bibinfo {volume} {638}},\ \bibinfo {pages} {L75} (\bibinfo {year}
  {2006}),\
  \Eprint{http://arxiv.org/abs/astro-ph/0511242}{arXiv:astro-ph/0511242}%
  \bibAnnoteFile{NoStop}{Hofloe06}%
\bibitem{GuaPorSip05}%
  \BibitemOpen
  \bibfield{author}{%
  \bibinfo {author} {\bibfnamefont{A.}~\bibnamefont{{Gualandris}}}, \bibinfo
  {author} {\bibfnamefont{S.}~\bibnamefont{{Portegies Zwart}}},\ and\ \bibinfo
  {author} {\bibfnamefont{M.~S.}\ \bibnamefont{{Sipior}}},\ }%
  \bibfield{journal}{%
  \Doi{10.1111/j.1365-2966.2005.09433.x}{\bibinfo {journal} {Mon. Not. R.
  Astron. Soc.}}\ }%
  \textbf{\bibinfo {volume} {363}},\ \bibinfo {pages} {223} (\bibinfo {month}
  {Oct.}\ \bibinfo {year} {2005}),\
  \Eprint{http://arxiv.org/abs/arXiv:astro-ph/0507365}{arXiv:astro-ph/0507365}%
  \bibAnnoteFile{NoStop}{GuaPorSip05}%
\bibitem{Mik83}%
  \BibitemOpen
  \bibfield{author}{%
  \bibinfo {author} {\bibfnamefont{S.}~\bibnamefont{{Mikkola}}},\ }%
  \bibfield{journal}{%
  \bibinfo {journal} {Mon. Not. R. Astron. Soc.}\ }%
  \textbf{\bibinfo {volume} {203}},\ \bibinfo {pages} {1107} (\bibinfo {month}
  {Jun.}\ \bibinfo {year} {1983}),\
  \url{http://adsabs.harvard.edu/abs/1983MNRAS.203.1107M}%
  \bibAnnoteFile{NoStop}{Mik83}%
\bibitem{Mik84}%
  \BibitemOpen
  \bibfield{author}{%
  \bibinfo {author} {\bibfnamefont{S.}~\bibnamefont{{Mikkola}}},\ }%
  \bibfield{journal}{%
  \bibinfo {journal} {Mon. Not. R. Astron. Soc.}\ }%
  \textbf{\bibinfo {volume} {207}},\ \bibinfo {pages} {115} (\bibinfo {month}
  {Mar.}\ \bibinfo {year} {1984}),\
  \url{http://adsabs.harvard.edu/abs/1984MNRAS.207..115M}%
  \bibAnnoteFile{NoStop}{Mik84}%
\bibitem{MilHam02}%
  \BibitemOpen
  \bibfield{author}{%
  \bibinfo {author} {\bibfnamefont{M.~C.}\ \bibnamefont{Miller}}\ and\ \bibinfo
  {author} {\bibfnamefont{D.~P.}\ \bibnamefont{Hamilton}},\ }%
  \bibfield{journal}{%
  \Doi{10.1086/341788}{\bibinfo {journal} {The Astrophysical Journal}}\ }%
  \textbf{\bibinfo {volume} {576}},\ \bibinfo {pages} {894} (\bibinfo {year}
  {2002}),\
  \Eprint{http://arxiv.org/abs/astro-ph/0202298}{arXiv:astro-ph/0202298}%
  \bibAnnoteFile{NoStop}{MilHam02}%
\bibitem{Hei01}%
  \BibitemOpen
  \bibfield{author}{%
  \bibinfo {author} {\bibfnamefont{P.}~\bibnamefont{{Hein{\"a}m{\"a}ki}}},\ }%
  \bibfield{journal}{%
  \Doi{10.1051/0004-6361:20010460}{\bibinfo {journal} {A\&A}}\ }%
  \textbf{\bibinfo {volume} {371}},\ \bibinfo {pages} {795} (\bibinfo {month}
  {Jun.}\ \bibinfo {year} {2001})%
  \bibAnnoteFile{NoStop}{Hei01}%
\bibitem{ValMik91}%
  \BibitemOpen
  \bibfield{author}{%
  \bibinfo {author} {\bibfnamefont{M.}~\bibnamefont{{Valtonen}}}\ and\ \bibinfo
  {author} {\bibfnamefont{S.}~\bibnamefont{{Mikkola}}},\ }%
  \bibfield{journal}{%
  \Doi{10.1146/annurev.aa.29.090191.000301}{\bibinfo {journal} {Annu. Rev.
  Astron. Astrophys.}}\ }%
  \textbf{\bibinfo {volume} {29}},\ \bibinfo {pages} {9} (\bibinfo {year}
  {1991})%
  \bibAnnoteFile{NoStop}{ValMik91}%
\bibitem{CamLouZlo07f}%
  \BibitemOpen
  \bibfield{author}{%
  \bibinfo {author} {\bibfnamefont{M.}~\bibnamefont{Campanelli}}, \bibinfo
  {author} {\bibfnamefont{C.~O.}\ \bibnamefont{Lousto}},\ and\ \bibinfo
  {author} {\bibfnamefont{Y.}~\bibnamefont{Zlochower}},\ }%
  \bibfield{journal}{%
  \Doi{10.1103/PhysRevD.77.101501}{\bibinfo {journal} {Phys. Rev. D}}\ }%
  \textbf{\bibinfo {volume} {77}},\ \bibinfo {pages} {101501(R)} (\bibinfo
  {year} {2008}),\ \Eprint{http://arxiv.org/abs/0710.0879}{arXiv:0710.0879
  [gr-qc]}%
  \bibAnnoteFile{NoStop}{CamLouZlo07f}%
\bibitem{LouZlo07a}%
  \BibitemOpen
  \bibfield{author}{%
  \bibinfo {author} {\bibfnamefont{C.~O.}\ \bibnamefont{Lousto}}\ and\ \bibinfo
  {author} {\bibfnamefont{Y.}~\bibnamefont{Zlochower}},\ }%
  \bibfield{journal}{%
  \Doi{10.1103/PhysRevD.77.024034}{\bibinfo {journal} {Phys. Rev. D}}\ }%
  \textbf{\bibinfo {volume} {77}},\ \bibinfo {pages} {024034} (\bibinfo {year}
  {2008}),\ \Eprint{http://arxiv.org/abs/0711.1165}{arXiv:0711.1165 [gr-qc]}%
  \bibAnnoteFile{NoStop}{LouZlo07a}%
\bibitem{GalBruCao10a}%
  \BibitemOpen
  \bibfield{author}{%
  \bibinfo {author} {\bibfnamefont{P.}~\bibnamefont{Galaviz}}, \bibinfo
  {author} {\bibfnamefont{B.}~\bibnamefont{Br{\"u}gmann}},\ and\ \bibinfo
  {author} {\bibfnamefont{Z.}~\bibnamefont{Cao}},\ }%
  \bibfield{journal}{%
  \Doi{10.1103/PhysRevD.82.024005}{\bibinfo {journal} {Phys. Rev. D}}\ }%
  \textbf{\bibinfo {volume} {82}},\ \bibinfo {pages} {024005} (\bibinfo {month}
  {Jul}\ \bibinfo {year} {2010}),\
  \Eprint{http://arxiv.org/abs/1004.1353}{arXiv:1004.1353 [gr-qc]}%
  \bibAnnoteFile{NoStop}{GalBruCao10a}%
\bibitem{Die03a}%
  \BibitemOpen
  \bibfield{author}{%
  \bibinfo {author} {\bibfnamefont{P.}~\bibnamefont{Diener}},\ }%
  \bibfield{journal}{%
  \Doi{10.1088/0264-9381/20/22/014}{\bibinfo {journal} {Class. Quantum Grav.}}\
  }%
  \textbf{\bibinfo {volume} {20}},\ \bibinfo {pages} {4901} (\bibinfo {year}
  {2003}),\ \Eprint{http://arxiv.org/abs/gr-qc/0305039}{arXiv:gr-qc/0305039}%
  \bibAnnoteFile{NoStop}{Die03a}%
\bibitem{JarLou10}%
  \BibitemOpen
  \bibfield{author}{%
  \bibinfo {author} {\bibfnamefont{G.}~\bibnamefont{{Jaramillo}}}\ and\
  \bibinfo {author} {\bibfnamefont{C.~O.}\ \bibnamefont{{Lousto}}},\ }%
  \bibfield{journal}{%
  \bibinfo {journal} {ArXiv e-prints}}%
   (\bibinfo {year} {2010}),\
  \Eprint{http://arxiv.org/abs/1008.2001}{arXiv:1008.2001 [gr-qc]}%
  \bibAnnoteFile{NoStop}{JarLou10}%
\bibitem{PonLouZlo10}%
  \BibitemOpen
  \bibfield{author}{%
  \bibinfo {author} {\bibfnamefont{M.}~\bibnamefont{{Ponce}}}, \bibinfo
  {author} {\bibfnamefont{C.}~\bibnamefont{{Lousto}}},\ and\ \bibinfo {author}
  {\bibfnamefont{Y.}~\bibnamefont{{Zlochower}}},\ }%
  \bibfield{journal}{%
  \bibinfo {journal} {ArXiv e-prints}}%
   (\bibinfo {year} {2010}),\
  \Eprint{http://arxiv.org/abs/1008.2761}{arXiv:1008.2761 [gr-qc]}%
  \bibAnnoteFile{NoStop}{PonLouZlo10}%
\bibitem{DamJarSch07a}%
  \BibitemOpen
  \bibfield{author}{%
  \bibinfo {author} {\bibfnamefont{T.}~\bibnamefont{Damour}}, \bibinfo {author}
  {\bibfnamefont{P.}~\bibnamefont{Jaranowski}},\ and\ \bibinfo {author}
  {\bibfnamefont{G.}~\bibnamefont{Sch{\"a}fer}},\ }%
  \bibfield{journal}{%
  \Doi{10.1103/PhysRevD.77.064032}{\bibinfo {journal} {Phys. Rev. D}}\ }%
  \textbf{\bibinfo {volume} {77}},\ \bibinfo {pages} {064032} (\bibinfo {year}
  {2008}),\ \Eprint{http://arxiv.org/abs/0711.1048}{arXiv:0711.1048 [gr-qc]}%
  \bibAnnoteFile{NoStop}{DamJarSch07a}%
\bibitem{SteHerSch08}%
  \BibitemOpen
  \bibfield{author}{%
  \bibinfo {author} {\bibfnamefont{J.}~\bibnamefont{Steinhoff}}, \bibinfo
  {author} {\bibfnamefont{S.}~\bibnamefont{Hergt}},\ and\ \bibinfo {author}
  {\bibfnamefont{G.}~\bibnamefont{Sch{\"a}fer}},\ }%
  \bibfield{journal}{%
  \Doi{10.1103/PhysRevD.77.081501}{\bibinfo {journal} {Phys. Rev. D}}\ }%
  \textbf{\bibinfo {volume} {77}},\ \bibinfo {pages} {081501} (\bibinfo {month}
  {Apr}\ \bibinfo {year} {2008}),\
  \Eprint{http://arxiv.org/abs/0712.1716}{arXiv:0712.1716 [gr-qc]}%
  \bibAnnoteFile{NoStop}{SteHerSch08}%
\bibitem{HarSte10}%
  \BibitemOpen
  \bibfield{author}{%
  \bibinfo {author} {\bibfnamefont{J.}~\bibnamefont{Hartung}}\ and\ \bibinfo
  {author} {\bibfnamefont{J.}~\bibnamefont{Steinhoff}},\ }%
  \bibfield{journal}{%
  \bibinfo {journal} {ArXiv e-prints}}%
   (\bibinfo {year} {2010}),\
  \Eprint{http://arxiv.org/abs/1011.1179}{arXiv:1011.1179 [gr-qc]}%
  \bibAnnoteFile{NoStop}{HarSte10}%
\bibitem{JarSch97}%
  \BibitemOpen
  \bibfield{author}{%
  \bibinfo {author} {\bibfnamefont{P.}~\bibnamefont{Jaranowski}}\ and\ \bibinfo
  {author} {\bibfnamefont{G.}~\bibnamefont{Sch{\"a}fer}},\ }%
  \bibfield{journal}{%
  \Doi{10.1103/PhysRevD.55.4712}{\bibinfo {journal} {Phys. Rev. D}}\ }%
  \textbf{\bibinfo {volume} {55}},\ \bibinfo {pages} {4712} (\bibinfo {year}
  {1997})%
  \bibAnnoteFile{NoStop}{JarSch97}%
\bibitem{Bla06a}%
  \BibitemOpen
  \bibfield{author}{%
  \bibinfo {author} {\bibfnamefont{L.}~\bibnamefont{Blanchet}},\ }%
  \bibfield{journal}{%
  \bibinfo {journal} {Living Reviews in Relativity}\ }%
  \textbf{\bibinfo {volume} {9}} (\bibinfo {year} {2006}),\
  \url{http://www.livingreviews.org/lrr-2006-4}%
  \bibAnnoteFile{NoStop}{Bla06a}%
\bibitem{FutIto07}%
  \BibitemOpen
  \bibfield{author}{%
  \bibinfo {author} {\bibfnamefont{T.}~\bibnamefont{Futamase}}\ and\ \bibinfo
  {author} {\bibfnamefont{Y.}~\bibnamefont{Itoh}},\ }%
  \bibfield{journal}{%
  \bibinfo {journal} {Living Reviews in Relativity}\ }%
  \textbf{\bibinfo {volume} {10}} (\bibinfo {year} {2007}),\
  \url{http://www.livingreviews.org/lrr-2007-2}%
  \bibAnnoteFile{NoStop}{FutIto07}%
\bibitem{DamJarSch00}%
  \BibitemOpen
  \bibfield{author}{%
  \bibinfo {author} {\bibfnamefont{T.}~\bibnamefont{Damour}}, \bibinfo {author}
  {\bibfnamefont{P.}~\bibnamefont{Jaranowski}},\ and\ \bibinfo {author}
  {\bibfnamefont{G.}~\bibnamefont{Sch{\"a}fer}},\ }%
  \bibfield{journal}{%
  \Doi{10.1103/PhysRevD.62.021501}{\bibinfo {journal} {Phys. Rev. D}}\ }%
  \textbf{\bibinfo {volume} {62}},\ \bibinfo {pages} {021501} (\bibinfo {year}
  {2000}),\ \Eprint{http://arxiv.org/abs/gr-qc/0003051}{arXiv:gr-qc/0003051}%
  \bibAnnoteFile{NoStop}{DamJarSch00}%
\bibitem{DamJarSch00a}%
  \BibitemOpen
  \bibfield{author}{%
  \bibinfo {author} {\bibfnamefont{T.}~\bibnamefont{Damour}}, \bibinfo {author}
  {\bibfnamefont{P.}~\bibnamefont{Jaranowski}},\ and\ \bibinfo {author}
  {\bibfnamefont{G.}~\bibnamefont{Sch{\"a}fer}},\ }%
  \bibfield{journal}{%
  \Doi{10.1103/PhysRevD.63.044021}{\bibinfo {journal} {Phys. Rev. D}}\ }%
  \textbf{\bibinfo {volume} {63}},\ \bibinfo {pages} {044021} (\bibinfo {year}
  {2001}),\ \Eprint{http://arxiv.org/abs/gr-qc/0010040}{arXiv:gr-qc/0010040}%
  \bibAnnoteFile{NoStop}{DamJarSch00a}%
\bibitem{DamJarSch01}%
  \BibitemOpen
  \bibfield{author}{%
  \bibinfo {author} {\bibfnamefont{T.}~\bibnamefont{Damour}}, \bibinfo {author}
  {\bibfnamefont{P.}~\bibnamefont{Jaranowski}},\ and\ \bibinfo {author}
  {\bibfnamefont{G.}~\bibnamefont{Sch{\"a}fer}},\ }%
  \bibfield{journal}{%
  \Doi{10.1016/S0370-2693(01)00642-6}{\bibinfo {journal} {Phys. Lett. B}}\ }%
  \textbf{\bibinfo {volume} {513}},\ \bibinfo {pages} {147} (\bibinfo {year}
  {2001}),\ \Eprint{http://arxiv.org/abs/gr-qc/0105038}{arXiv:gr-qc/0105038}%
  \bibAnnoteFile{NoStop}{DamJarSch01}%
\bibitem{BlaDamEsp04a}%
  \BibitemOpen
  \bibfield{author}{%
  \bibinfo {author} {\bibfnamefont{L.}~\bibnamefont{Blanchet}}, \bibinfo
  {author} {\bibfnamefont{T.}~\bibnamefont{Damour}},\ and\ \bibinfo {author}
  {\bibfnamefont{G.}~\bibnamefont{Esposito-Far{\`e}se}},\ }%
  \bibfield{journal}{%
  \Doi{10.1103/PhysRevD.69.124007}{\bibinfo {journal} {Phys. Rev. D}}\ }%
  \textbf{\bibinfo {volume} {69}},\ \bibinfo {pages} {124007} (\bibinfo {month}
  {Jun}\ \bibinfo {year} {2004}),\
  \Eprint{http://arxiv.org/abs/gr-qc/0311052}{arXiv:gr-qc/0311052}%
  \bibAnnoteFile{NoStop}{BlaDamEsp04a}%
\bibitem{KonFaySch03}%
  \BibitemOpen
  \bibfield{author}{%
  \bibinfo {author} {\bibfnamefont{C.}~\bibnamefont{K{\"o}nigsd{\"o}rffer}},
  \bibinfo {author} {\bibfnamefont{G.}~\bibnamefont{Faye}},\ and\ \bibinfo
  {author} {\bibfnamefont{G.}~\bibnamefont{Sch{\"a}fer}},\ }%
  \bibfield{journal}{%
  \Doi{10.1103/PhysRevD.68.044004}{\bibinfo {journal} {Phys. Rev. D}}\ }%
  \textbf{\bibinfo {volume} {68}},\ \bibinfo {pages} {044004} (\bibinfo {year}
  {2003})%
  \bibAnnoteFile{NoStop}{KonFaySch03}%
\bibitem{Sch87}%
  \BibitemOpen
  \bibfield{author}{%
  \bibinfo {author} {\bibfnamefont{G.}~\bibnamefont{Sch{\"a}fer}},\ }%
  \bibfield{journal}{%
  \Doi{10.1016/0375-9601(87)90389-6}{\bibinfo {journal} {Physics Letters A}}\
  }%
  \textbf{\bibinfo {volume} {123}},\ \bibinfo {pages} {336} (\bibinfo {year}
  {1987})%
  \bibAnnoteFile{NoStop}{Sch87}%
\bibitem{LouNak08}%
  \BibitemOpen
  \bibfield{author}{%
  \bibinfo {author} {\bibfnamefont{C.~O.}\ \bibnamefont{Lousto}}\ and\ \bibinfo
  {author} {\bibfnamefont{H.}~\bibnamefont{Nakano}},\ }%
  \bibfield{journal}{%
  \Doi{10.1088/0264-9381/25/19/195019}{\bibinfo {journal} {Class. Quantum
  Grav.}}\ }%
  \textbf{\bibinfo {volume} {25}},\ \bibinfo {pages} {195019} (\bibinfo {year}
  {2008}),\ \Eprint{http://arxiv.org/abs/0710.5542}{arXiv:0710.5542 [gr-qc]}%
  \bibAnnoteFile{NoStop}{LouNak08}%
\bibitem{Chu09}%
  \BibitemOpen
  \bibfield{author}{%
  \bibinfo {author} {\bibfnamefont{Y.}~\bibnamefont{Chu}},\ }%
  \bibfield{journal}{%
  \Doi{10.1103/PhysRevD.79.044031}{\bibinfo {journal} {Phys. Rev. D}}\ }%
  \textbf{\bibinfo {volume} {79}},\ \bibinfo {pages} {044031} (\bibinfo {month}
  {Feb}\ \bibinfo {year} {2009}),\
  \Eprint{http://arxiv.org/abs/0812.0012}{arXiv:0812.0012 [gr-qc]}%
  \bibAnnoteFile{NoStop}{Chu09}%
\bibitem{OhtKimHii75}%
  \BibitemOpen
  \bibfield{author}{%
  \bibinfo {author} {\bibfnamefont{T.}~\bibnamefont{Ohta}}, \bibinfo {author}
  {\bibfnamefont{T.}~\bibnamefont{Kimura}},\ and\ \bibinfo {author}
  {\bibfnamefont{K.}~\bibnamefont{Hiida}},\ }%
  \bibfield{journal}{%
  \Doi{10.1007/BF02726342}{\bibinfo {journal} {Nuovo Cim.}}\ }%
  \textbf{\bibinfo {volume} {B27}},\ \bibinfo {pages} {103} (\bibinfo {year}
  {1975})%
  \bibAnnoteFile{NoStop}{OhtKimHii75}%
\bibitem{MitWil07}%
  \BibitemOpen
  \bibfield{author}{%
  \bibinfo {author} {\bibfnamefont{T.}~\bibnamefont{Mitchell}}\ and\ \bibinfo
  {author} {\bibfnamefont{C.~M.}\ \bibnamefont{Will}},\ }%
  \bibfield{journal}{%
  \Doi{10.1103/PhysRevD.75.124025}{\bibinfo {journal} {Phys. Rev. D}}\ }%
  \textbf{\bibinfo {volume} {75}},\ \bibinfo {pages} {124025} (\bibinfo {month}
  {Jun}\ \bibinfo {year} {2007}),\
  \Eprint{http://arxiv.org/abs/0704.2243}{arXiv:0704.2243 [gr-qc]}%
  \bibAnnoteFile{NoStop}{MitWil07}%
\bibitem{Moo93}%
  \BibitemOpen
  \bibfield{author}{%
  \bibinfo {author} {\bibfnamefont{C.}~\bibnamefont{Moore}},\ }%
  \bibfield{journal}{%
  \Doi{10.1103/PhysRevLett.70.3675}{\bibinfo {journal} {Phys. Rev. Lett.}}\ }%
  \textbf{\bibinfo {volume} {70}},\ \bibinfo {pages} {3675} (\bibinfo {month}
  {Jun}\ \bibinfo {year} {1993})%
  \bibAnnoteFile{NoStop}{Moo93}%
\bibitem{TatTakHid07}%
  \BibitemOpen
  \bibfield{author}{%
  \bibinfo {author} {\bibfnamefont{T.}~\bibnamefont{Imai}}, \bibinfo {author}
  {\bibfnamefont{T.}~\bibnamefont{Chiba}},\ and\ \bibinfo {author}
  {\bibfnamefont{H.}~\bibnamefont{Asada}},\ }%
  \bibfield{journal}{%
  \Doi{10.1103/PhysRevLett.98.201102}{\bibinfo {journal} {Phys. Rev. Lett.}}\
  }%
  \textbf{\bibinfo {volume} {98}},\ \bibinfo {pages} {201102} (\bibinfo {month}
  {May}\ \bibinfo {year} {2007})%
  \bibAnnoteFile{NoStop}{TatTakHid07}%
\bibitem{YamAsa10}%
  \BibitemOpen
  \bibfield{author}{%
  \bibinfo {author} {\bibfnamefont{K.}~\bibnamefont{Yamada}}\ and\ \bibinfo
  {author} {\bibfnamefont{H.}~\bibnamefont{Asada}},\ }%
  \bibfield{journal}{%
  \Doi{10.1103/PhysRevD.82.104019}{\bibinfo {journal} {Phys. Rev. D}}\ }%
  \textbf{\bibinfo {volume} {82}},\ \bibinfo {pages} {104019} (\bibinfo {month}
  {Nov}\ \bibinfo {year} {2010}),\
  \Eprint{http://arxiv.org/abs/1010.2284}{arXiv:1010.2284 [gr-qc]}%
  \bibAnnoteFile{NoStop}{YamAsa10}%
\bibitem{YamAsa10a}%
  \BibitemOpen
  \bibfield{author}{%
  \bibinfo {author} {\bibfnamefont{K.}~\bibnamefont{Yamada}}\ and\ \bibinfo
  {author} {\bibfnamefont{H.}~\bibnamefont{Asada}},\ }%
  \bibfield{journal}{%
  \bibinfo {journal} {ArXiv e-prints}}%
   (\bibinfo {year} {2010}),\
  \Eprint{http://arxiv.org/abs/1011.2007}{arXiv:1011.2007 [gr-qc]}%
  \bibAnnoteFile{NoStop}{YamAsa10a}%
\bibitem{IchYamAsa10}%
  \BibitemOpen
  \bibfield{author}{%
  \bibinfo {author} {\bibfnamefont{T.}~\bibnamefont{Ichita}}, \bibinfo {author}
  {\bibfnamefont{K.}~\bibnamefont{Yamada}},\ and\ \bibinfo {author}
  {\bibfnamefont{H.}~\bibnamefont{Asada}},\ }%
  \bibfield{journal}{%
  \bibinfo {journal} {ArXiv e-prints}}%
   (\bibinfo {year} {2010}),\
  \Eprint{http://arxiv.org/abs/1011.3886}{arXiv:1011.3886 [gr-qc]}%
  \bibAnnoteFile{NoStop}{IchYamAsa10}%
\bibitem{JSch10}%
  \BibitemOpen
  \bibfield{author}{%
  \bibinfo {author} {\bibfnamefont{J.~D.}\ \bibnamefont{Schnittman}},\ }%
  \bibfield{journal}{%
  \bibinfo {journal} {ArXiv e-prints}}%
   (\bibinfo {year} {2010}),\
  \Eprint{http://arxiv.org/abs/1006.0182}{arXiv:1006.0182 [astro-ph.HE]}%
  \bibAnnoteFile{NoStop}{JSch10}%
\bibitem{AmaDew10}%
  \BibitemOpen
  \bibfield{author}{%
  \bibinfo {author} {\bibfnamefont{P.}~\bibnamefont{{Amaro-Seoane}}}\ and\
  \bibinfo {author} {\bibfnamefont{M.}~\bibnamefont{{Dewi Freitag}}},\ }%
  \bibfield{journal}{%
  \bibinfo {journal} {ArXiv e-prints}}%
   (\bibinfo {year} {2010}),\
  \Eprint{http://arxiv.org/abs/1009.1870}{arXiv:1009.1870 [astro-ph.CO]}%
  \bibAnnoteFile{NoStop}{AmaDew10}%
\bibitem{LevCon10}%
  \BibitemOpen
  \bibfield{author}{%
  \bibinfo {author} {\bibfnamefont{J.}~\bibnamefont{{Levin}}}\ and\ \bibinfo
  {author} {\bibfnamefont{H.}~\bibnamefont{{Contreras}}},\ }%
  \bibfield{journal}{%
  \bibinfo {journal} {ArXiv e-prints}}%
   (\bibinfo {month} {Sep.}\ \bibinfo {year} {2010}),\
  \Eprint{http://arxiv.org/abs/1009.2533}{arXiv:1009.2533 [gr-qc]}%
  \bibAnnoteFile{NoStop}{LevCon10}%
\bibitem{Szi06}%
  \BibitemOpen
  \bibfield{author}{%
  \bibinfo {author} {\bibfnamefont{T.}~\bibnamefont{Szirtes}},\ }%
  \emph{\bibinfo {title} {Applied Dimensional Analysis and Modeling}},\
  \bibinfo {edition} {2nd}\ ed.\ (\bibinfo {publisher}
  {Butterworth-Heinemann},\ \bibinfo {year} {2006})%
  \bibAnnoteFile{NoStop}{Szi06}%
\bibitem{Mag07a}%
  \BibitemOpen
  \bibfield{author}{%
  \bibinfo {author} {\bibfnamefont{M.}~\bibnamefont{Maggiore}},\ }%
  \emph{\bibinfo {title} {Gravitational Waves}},\ Vol.~\bibinfo {volume} {1}\
  (\bibinfo {publisher} {Oxford University Press, USA.},\ \bibinfo {year}
  {2007})%
  \bibAnnoteFile{NoStop}{Mag07a}%
\bibitem{DamSch85}%
  \BibitemOpen
  \bibfield{author}{%
  \bibinfo {author} {\bibfnamefont{T.}~\bibnamefont{Damour}}\ and\ \bibinfo
  {author} {\bibfnamefont{G.}~\bibnamefont{Sch{\"a}fer}},\ }%
  \bibfield{journal}{%
  \bibinfo {journal} {Gen. Rel. Grav.}\ }%
  \textbf{\bibinfo {volume} {17}},\ \bibinfo {pages} {879} (\bibinfo {year}
  {1985}),\ \url{http://www.springerlink.com/content/v1j0847845641337/}%
  \bibAnnoteFile{NoStop}{DamSch85}%
\bibitem{FlaHug05}%
  \BibitemOpen
  \bibfield{author}{%
  \bibinfo {author} {\bibfnamefont{E.~E.}\ \bibnamefont{Flanagan}}\ and\
  \bibinfo {author} {\bibfnamefont{S.~A.}\ \bibnamefont{Hughes}},\ }%
  \bibfield{journal}{%
  \Doi{10.1088/1367-2630/7/1/204}{\bibinfo {journal} {New J. Phys.}}\ }%
  \textbf{\bibinfo {volume} {7}},\ \bibinfo {pages} {204} (\bibinfo {year}
  {2005}),\ \Eprint{http://arxiv.org/abs/gr-qc/0501041}{arXiv:gr-qc/0501041}%
  \bibAnnoteFile{NoStop}{FlaHug05}%
\bibitem{Wol08}%
  \BibitemOpen
  \bibfield{author}{%
  \bibinfo {author} {\bibnamefont{{Wolfram Research, Inc.}}},\ }%
  \emph{\bibinfo {title} {Mathematica}},\ \bibinfo {edition} {version 7.0}\
  ed.\ (\bibinfo {publisher} {{Wolfram Research, Inc.}},\ \bibinfo {year}
  {2008})%
  \bibAnnoteFile{NoStop}{Wol08}%
\bibitem{Bur67}%
  \BibitemOpen
  \bibfield{author}{%
  \bibinfo {author} {\bibfnamefont{C.~A.}\ \bibnamefont{{Burdet}}},\ }%
  \bibfield{journal}{%
  \Doi{10.1007/BF01601283}{\bibinfo {journal} {Zeitschrift Angewandte
  Mathematik und Physik}}\ }%
  \textbf{\bibinfo {volume} {18}},\ \bibinfo {pages} {434} (\bibinfo {month}
  {May}\ \bibinfo {year} {1967})%
  \bibAnnoteFile{NoStop}{Bur67}%
\bibitem{Heg74}%
  \BibitemOpen
  \bibfield{author}{%
  \bibinfo {author} {\bibfnamefont{D.~C.}\ \bibnamefont{{Heggie}}},\ }%
  \bibfield{journal}{%
  \Doi{10.1007/BF01227621}{\bibinfo {journal} {Celestial Mechanics and
  Dynamical Astronomy}}\ }%
  \textbf{\bibinfo {volume} {10}},\ \bibinfo {pages} {217} (\bibinfo {month}
  {Oct.}\ \bibinfo {year} {1974})%
  \bibAnnoteFile{NoStop}{Heg74}%
\bibitem{MikAar89}%
  \BibitemOpen
  \bibfield{author}{%
  \bibinfo {author} {\bibfnamefont{S.}~\bibnamefont{{Mikkola}}}\ and\ \bibinfo
  {author} {\bibfnamefont{S.~J.}\ \bibnamefont{{Aarseth}}},\ }%
  \bibfield{journal}{%
  \Doi{10.1007/BF00051012}{\bibinfo {journal} {Celestial Mechanics and
  Dynamical Astronomy}}\ }%
  \textbf{\bibinfo {volume} {47}},\ \bibinfo {pages} {375} (\bibinfo {month}
  {Dec.}\ \bibinfo {year} {1989})%
  \bibAnnoteFile{NoStop}{MikAar89}%
\bibitem{MikAar93}%
  \BibitemOpen
  \bibfield{author}{%
  \bibinfo {author} {\bibfnamefont{S.}~\bibnamefont{{Mikkola}}}\ and\ \bibinfo
  {author} {\bibfnamefont{S.~J.}\ \bibnamefont{{Aarseth}}},\ }%
  \bibfield{journal}{%
  \Doi{10.1007/BF00695714}{\bibinfo {journal} {Celestial Mechanics and
  Dynamical Astronomy}}\ }%
  \textbf{\bibinfo {volume} {57}},\ \bibinfo {pages} {439} (\bibinfo {month}
  {Nov.}\ \bibinfo {year} {1993})%
  \bibAnnoteFile{NoStop}{MikAar93}%
\bibitem{MikAar96}%
  \BibitemOpen
  \bibfield{author}{%
  \bibinfo {author} {\bibfnamefont{S.}~\bibnamefont{{Mikkola}}}\ and\ \bibinfo
  {author} {\bibfnamefont{S.~J.}\ \bibnamefont{{Aarseth}}},\ }%
  \bibfield{journal}{%
  \Doi{10.1007/BF00728347}{\bibinfo {journal} {Celestial Mechanics and
  Dynamical Astronomy}}\ }%
  \textbf{\bibinfo {volume} {64}},\ \bibinfo {pages} {197} (\bibinfo {month}
  {Sep.}\ \bibinfo {year} {1996})%
  \bibAnnoteFile{NoStop}{MikAar96}%
\bibitem{GolPooSaf01}%
  \BibitemOpen
  \bibfield{author}{%
  \bibinfo {author} {\bibfnamefont{H.}~\bibnamefont{Goldstein}}, \bibinfo
  {author} {\bibfnamefont{C.~P.}\ \bibnamefont{Poole}},\ and\ \bibinfo {author}
  {\bibfnamefont{J.}~\bibnamefont{Safko}},\ }%
  \emph{\bibinfo {title} {Classical Mechanics}}\ (\bibinfo {publisher} {Addison
  Wesley},\ \bibinfo {year} {2001})\ ISBN \bibinfo {isbn} {0-201-65702-3}%
  \bibAnnoteFile{NoStop}{GolPooSaf01}%
\bibitem{Pet64}%
  \BibitemOpen
  \bibfield{author}{%
  \bibinfo {author} {\bibfnamefont{P.~C.}\ \bibnamefont{Peters}},\ }%
  \bibfield{journal}{%
  \Doi{10.1103/PhysRev.136.B1224}{\bibinfo {journal} {Phys. Rev.}}\ }%
  \textbf{\bibinfo {volume} {136}},\ \bibinfo {pages} {B1224} (\bibinfo {year}
  {1964})%
  \bibAnnoteFile{NoStop}{Pet64}%
\bibitem{Hen76}%
  \BibitemOpen
  \bibfield{author}{%
  \bibinfo {author} {\bibfnamefont{M.}~\bibnamefont{H{\'e}non}},\ }%
  \bibfield{journal}{%
  \Doi{10.1007/BF01228647}{\bibinfo {journal} {Celestial Mechanics and
  Dynamical Astronomy}}\ }%
  \textbf{\bibinfo {volume} {13}},\ \bibinfo {pages} {267} (\bibinfo {year}
  {1976})%
  \bibAnnoteFile{NoStop}{Hen76}%
\bibitem{Nau01}%
  \BibitemOpen
  \bibfield{author}{%
  \bibinfo {author} {\bibfnamefont{M.}~\bibnamefont{{Nauenberg}}},\ }%
  \bibfield{journal}{%
  \Doi{10.1016/S0375-9601(01)00768-X}{\bibinfo {journal} {Physics Letters A}}\
  }%
  \textbf{\bibinfo {volume} {292}},\ \bibinfo {pages} {93} (\bibinfo {month}
  {Dec.}\ \bibinfo {year} {2001}),\
  \Eprint{http://arxiv.org/abs/arXiv:nlin/0112003v2}{arXiv:nlin/0112003v2}%
  \bibAnnoteFile{NoStop}{Nau01}%
\bibitem{MooNau08}%
  \BibitemOpen
  \bibfield{author}{%
  \bibinfo {author} {\bibfnamefont{C.}~\bibnamefont{{Moore}}}\ and\ \bibinfo
  {author} {\bibfnamefont{M.}~\bibnamefont{{Nauenberg}}},\ }%
  \bibfield{journal}{%
  \bibinfo {journal} {ArXiv e-prints}}%
   (\bibinfo {year} {2008}),\
  \Eprint{http://arxiv.org/abs/math/0511219v2}{arXiv:math/0511219v2}%
  \bibAnnoteFile{NoStop}{MooNau08}%
\bibitem{Lev00}%
  \BibitemOpen
  \bibfield{author}{%
  \bibinfo {author} {\bibfnamefont{J.}~\bibnamefont{{Levin}}},\ }%
  \bibfield{journal}{%
  \Doi{10.1103/PhysRevLett.84.3515}{\bibinfo {journal} {Phys. Rev. D}}\ }%
  \textbf{\bibinfo {volume} {84}},\ \bibinfo {pages} {3515} (\bibinfo {month}
  {Apr.}\ \bibinfo {year} {2000}),\
  \Eprint{http://arxiv.org/abs/arXiv:gr-qc/9910040}{arXiv:gr-qc/9910040}%
  \bibAnnoteFile{NoStop}{Lev00}%
\bibitem{CorLev03}%
  \BibitemOpen
  \bibfield{author}{%
  \bibinfo {author} {\bibfnamefont{N.~J.}\ \bibnamefont{{Cornish}}}\ and\
  \bibinfo {author} {\bibfnamefont{J.}~\bibnamefont{{Levin}}},\ }%
  \bibfield{journal}{%
  \Doi{10.1103/PhysRevD.68.024004}{\bibinfo {journal} {Phys. Rev. D}}\ }%
  \textbf{\bibinfo {volume} {68}},\ \bibinfo {pages} {024004} (\bibinfo {month}
  {Jul.}\ \bibinfo {year} {2003})%
  \bibAnnoteFile{NoStop}{CorLev03}%
\bibitem{BarLev03}%
  \BibitemOpen
  \bibfield{author}{%
  \bibinfo {author} {\bibfnamefont{J.~D.}\ \bibnamefont{{Barrow}}}\ and\
  \bibinfo {author} {\bibfnamefont{J.}~\bibnamefont{{Levin}}}}%
   (\bibinfo {month} {Mar.}\ \bibinfo {year} {2003}),\
  \Eprint{http://arxiv.org/abs/nlin/0303070}{arXiv:nlin/0303070 [nlin.CD]}%
  \bibAnnoteFile{NoStop}{BarLev03}%
\bibitem{Lev06}%
  \BibitemOpen
  \bibfield{author}{%
  \bibinfo {author} {\bibfnamefont{J.}~\bibnamefont{{Levin}}},\ }%
  \bibfield{journal}{%
  \Doi{10.1103/PhysRevD.74.124027}{\bibinfo {journal} {Phys. Rev. D}}\ }%
  \textbf{\bibinfo {volume} {74}},\ \bibinfo {pages} {124027} (\bibinfo {month}
  {Dec.}\ \bibinfo {year} {2006}),\
  \Eprint{http://arxiv.org/abs/arXiv:gr-qc/0612003}{arXiv:gr-qc/0612003}%
  \bibAnnoteFile{NoStop}{Lev06}%
\bibitem{GopKni05}%
  \BibitemOpen
  \bibfield{author}{%
  \bibinfo {author} {\bibfnamefont{A.}~\bibnamefont{{Gopakumar}}}\ and\
  \bibinfo {author} {\bibfnamefont{C.}~\bibnamefont{{K{\"o}nigsd{\"o}rffer}}},\
  }%
  \bibfield{journal}{%
  \Doi{10.1103/PhysRevD.72.121501}{\bibinfo {journal} {Phys. Rev. D}}\ }%
  \textbf{\bibinfo {volume} {72}},\ \bibinfo {pages} {121501} (\bibinfo {month}
  {Dec.}\ \bibinfo {year} {2005}),\
  \Eprint{http://arxiv.org/abs/arXiv:gr-qc/0511009}{arXiv:gr-qc/0511009}%
  \bibAnnoteFile{NoStop}{GopKni05}%
\bibitem{Han08}%
  \BibitemOpen
  \bibfield{author}{%
  \bibinfo {author} {\bibfnamefont{W.}~\bibnamefont{{Han}}},\ }%
  \bibfield{journal}{%
  \Doi{10.1007/s10714-007-0598-9}{\bibinfo {journal} {Gen. Rel. Grav.}}\ }%
  \textbf{\bibinfo {volume} {40}},\ \bibinfo {pages} {1831} (\bibinfo {month}
  {Sep.}\ \bibinfo {year} {2008}),\
  \Eprint{http://arxiv.org/abs/1006.2229}{arXiv:1006.2229 [gr-qc]}%
  \bibAnnoteFile{NoStop}{Han08}%
\bibitem{SchRas01}%
  \BibitemOpen
  \bibfield{author}{%
  \bibinfo {author} {\bibfnamefont{J.~D.}\ \bibnamefont{{Schnittman}}}\ and\
  \bibinfo {author} {\bibfnamefont{F.~A.}\ \bibnamefont{{Rasio}}},\ }%
  \bibfield{journal}{%
  \Doi{10.1103/PhysRevLett.87.121101}{\bibinfo {journal} {Phys. Rev. Lett}}\ }%
  \textbf{\bibinfo {volume} {87}},\ \bibinfo {pages} {121101} (\bibinfo {month}
  {Sep.}\ \bibinfo {year} {2001}),\
  \Eprint{http://arxiv.org/abs/arXiv:gr-qc/0107082}{arXiv:gr-qc/0107082}%
  \bibAnnoteFile{NoStop}{SchRas01}%
\end{thebibliography}%


\end{document}